\documentclass[prx,twocolumn,showpacs]{revtex4-1}
\usepackage{ae}
\usepackage{amsmath}
\usepackage{amsfonts}
\usepackage{fontenc}
\usepackage{graphicx}
\usepackage{epstopdf}
\usepackage{fancyhdr}
\usepackage{color}
\usepackage{xcolor}
\usepackage{float}
\usepackage{hyperref}
\usepackage{tikz}
\usepackage{dcolumn}
\usepackage{bm}
\usepackage{lipsum}
\usepackage{lineno}

\makeatletter
\newcommand*\bigcdot{\mathpalette\bigcdot@{.5}}
\newcommand*\bigcdot@[2]{\mathbin{\vcenter{\hbox{\scalebox{#2}{$\m@th#1\bullet$}}}}}
\makeatother

\newcommand{\medio}[1]{\left\langle #1\right\rangle}
\addtocounter{secnumdepth}{1}

\begin{document}
\title{First-passage distributions for the one--dimensional Fokker-Planck equation}
\author{Oriol Artime, Nagi Khalil, Ra\'ul Toral, and Maxi San Miguel}
\affiliation{IFISC (CSIC-UIB), Instituto de F\'isica Interdisciplinar y Sistemas Complejos, Campus Universitat de les Illes Balears, E-07122 Palma de Mallorca, Spain}
\date{\today}

\begin{abstract}
 We present an analytical framework to study the first-passage (FP) and first-return (FR) distributions for the broad family of models described by the one-dimensional Fokker-Planck equation in finite domains, identifying general properties of these distributions for different classes of models.  When in the Fokker-Planck equation the diffusion coefficient is positive (nonzero) and the drift term is bounded, as in the case of a Brownian walker, both distributions may exhibit a power-law decay with exponent $-3/2$ for intermediate times. We discuss how the influence of an absorbing state changes this exponent. The absorbing state is characterized by a vanishing diffusion coefficient and/or a diverging drift term. Remarkably, the exponent of the Brownian walker class of models is still found, as long as the departure and arrival regions are far enough from the absorbing state, but the range of times where the power law is observed narrows. Close enough to the absorbing point, though, a new exponent may appear. The particular value of the exponent depends on the behavior of the diffusion and the drift terms of the Fokker-Planck equation. We focus on the case of a diffusion term vanishing linearly at the absorbing point. In this case, the FP and FR distributions are similar to those of the voter model, characterized by a power law with exponent $-2$. As an illustration of the general theory, we compare it with exact analytical solutions and extensive numerical simulations of a two-parameter voter-like family models. We study the behavior of the FP and FR distributions by tuning the importance of the absorbing points throughout changes of the parameters. Finally, the possibility of inferring relevant information about the steady-sate probability distribution of a model from the FP and FR distributions is addressed.
\end{abstract}
\pacs{02.50.Ey,05.10.Gg,89.75.-k}

\maketitle

\section{Introduction\label{sec:1}}

The first-passage (FP) and first-return (FR) times of a stochastic variable are defined as the times needed for the variable to reach a particular value from a given initial condition  or to return to the initial condition for the first time. They provide valuable information on the temporal properties of the system and in turn are relatively easy to obtain experimentally or by means of numerical simulations \cite{fe71,we94,re01,ga09,meosre14}. As consequence, the calculation of these quantities  has had immediate applicability in a myriad of problems: spreading of diseases \cite{llma01}, animal or human movement \cite{gohiba08}, neuron firing dynamics \cite{tu05}, diffusion in bounded domains \cite{cobemo05,cobetevokl07,gr07,rubegrvo12,bevo14,gr15}, diffusion in expanding mediums \cite{yuabes16}, diffusion-controlled reactions \cite{szscsc80}, controlled kinetics \cite{bechklmevo10,gome16a}, the computation of reaction rates in chemical reactions \cite{karebe10}, run-and-tumble particles \cite{majekukusamaredh18}, renewal and nonrenewal systems \cite{pt18}, or nonequilibrium systems in general \cite{brmasc13}.

One of the most simple problems where first-passage properties have been studied is the random walk on a semi-infinite line. The position $x_n$ of the walker after $n$ time steps verifies $x_n=x_{n-1}+\eta_n$ where the jumps $\eta_n$ are identically and independent random variables. Here the interest is in the persistence probability $Q(x_0,n)$ or the probability of the random walk with the initial condition $x=x_0$ to survive (not to reach the origin) until at least time $n$. From the persistence probability one can obtain the FP probability to reach the origin at time $n$ starting at $x_0>0$ as, $f(x=0,n|x_0)=Q(x_0,n)-Q(x_0,n+1)$. The same formula holds, with $x_0=0$, for the FR probability to return to the origin at time $n$.  Sparre-Andersen \cite{an53} showed that $Q(0,n)$ (and hence the FR to the origin) has a universal character, meaning that $Q(0,n)=\left(\begin{array}{c} 2n \\ n\end{array}\right)2^{-2n}$ for any symmetric and continuous jump distribution. This implies an algebraic decay of $Q(0,n)$ as $n^{-1/2}$ and of $P(0,n)$ as $n^{-3/2}$ for $n\rightarrow\infty$. The study of persistence exponents has been also extended to a random walk in a lattice \cite{we94,mascwe12} with similar results, and to non-symmetric jump distributions \cite{an54,mascwe12}. In the latter case, one of the main results is the breakdown of the aforementioned universality; in particular, if the jump distribution has an infinite second moment, the decay of the persistence probability is not purely algebraic (see \cite{mascwe12,brmasc13}).

Another relevant example is the Brownian walker, a continuous-time version of the random walker \cite{re01}. In this case, the stochastic process becomes a purely diffusive process, so the probability distribution for the position of the walker satisfies a Fokker-Plank equation with a constant diffusion coefficient and zero drift term. In the one-dimensional (1D) case the FP distribution is calculated \cite{ka07} and it decays in time as a power law with exponent $-3/2$, but now with an exponential cutoff if the domain is bounded \cite{re01}. Within the context of the 1D Fokker-Planck equation, other decay exponents, different from the one of the Brownian walker, have been observed for the FP and FR distributions. This is the case of neural avalanches \cite{mahimasaplmu17}, described by a Fokker-Planck equation with a position-dependent diffusion coefficient, where the FP distribution to the state of no active neurons from an infinitesimally close active state shows a power law of exponent $-2$. An important issue to take into account when explaining such a decay is the absorbing nature of the boundary.
In other systems with absorbing states, such as the voter model \cite{li12}, dynamic percolation \cite{gr83} or the Manna model \cite{vedimuza98}, the FP distribution, from a departure point very close to the boundary towards the absorbing state, also shows an exponent $-2$ \cite{savibumu17}.

In this work we explore the FP and FR time distributions of the family of models described by a 1D Fokker-Planck equation in finite domains, with state-dependent drift and diffusion terms. These models can be seen as the continuous-time limit of the random walker with an $x$-dependent nonsymmetric jump distribution with finite second moment. In the language of the random walker, if the length of the system explored by the walker is large enough, we provide compelling evidence, using a general approximate methodology, that the FP and FR distributions have an intermediate power-law time decay whose exponents can take different values. The length explored by the walker depends on the system size and on the location of the initial and final (target) positions and the values of the exponents depend on some conditions involving the diffusion and drift terms as well as the initial and final positions. For the class of models for which the diffusion does not vanish and the drift remains bounded, the exponent is $-3/2$, hence the result of Sparre-Andersen is generally obtained. Furthermore, the exponent $-2$ may appear when the diffusion coefficient vanishes linearly at some accessible position of the random walk. Observe, however, that the intermediate power-law time decays of both cases can be followed by a non-universal scaling which also might have a power-law scaling in time \cite{gome16}. 

As illustrations of the predictions of our general theory, we systematically study, both analytically and by extensive simulations, different models of increasing complexity. The models can be mapped into a family of two-parameter voter models which allow us to analyze the interplay between the $-3/2$ and $-2$ exponents. These models include the random walk, the Ornstein-Uhlenbeck process, the voter model itself \cite{clsu73,holi75,suegsa05,cafolo09,karebe10}, and two noisy variations \cite{ki93,grma95,khsato18,catosa16,pecasato18} of the latter.  We note, however, that the 1D  $x$ variable of the Fokker-Planck equation has different meanings for different models. While it represents a space variable for the random walk, it is a macroscopic density variable in the mean field approximation for voter models in large lattices of different spatial dimensionality.

The outline of the paper is as follows. In Sec.~\ref{sec:2} we briefly review the relation between the 1D Fokker-Planck equation and the FP distribution, leaving a more general treatment to Appendixes~\ref{app:1} and \ref{app:2} where we discuss in particular the difficulties of computing the FR distribution. In Sec.~\ref{sec:3} the FP distribution for the 1D Fokker-Planck equation is analyzed in detail. Among other features, we identify the conditions under which power laws with exponents $-3/2$ and $-2$ can appear in the FP distribution. The general theory of Sec.~\ref{sec:3} is tested in Sec.~\ref{sec:4}, where a family of voterlike models is studied in depth, theoretically and by means of Monte Carlo simulations. Finally, Sec. \ref{sec:5} contains a summary and our conclusions.

\section{From the 1D Fokker-Planck equation to the first-passage distribution\label{sec:2}}
Consider a one-dimensional real stochastic variable $X(t)\in I$, with $I$ a closed interval. Its probability density $p(x,t)$ satisfies the Fokker-Planck equation,
\begin{equation}
 \label{eq:14}
 \frac{\partial p}{\partial t}=-\frac{\partial[A(x) p]}{\partial x}+\frac{1}{2}\frac{\partial^2[B(x) p]}{\partial x^2}\equiv -\frac{\partial J[x|p]}{\partial x},
\end{equation}
where $A(x)$ and $B(x)\ge0$ are generic time-independent drift and diffusion coefficients and the probability flow $J[x|p]$, defined through the last equality, is a function of $x$ and a functional of the probability density $p$. Throughout this article we will be concerned with dynamics that are strictly governed by Eq.~\eqref{eq:14}, thus excluding some higher-dimensional problems that can be effectively reduced to one-dimensional equations but cannot be written in the form of the above equation. We call absorbing states of the dynamics those states in which once the system reaches them, it cannot leave. They can be interpreted in terms of the drift and diffusion coefficients of the Fokker-Planck equation: a point is absorbing if the diffusion coefficient vanishes and the drift is either null or points towards that state, or regardless of the diffusion, the drift coefficient is diverging and points towards the absorbing state. Our first objective is to compute the FP distribution $f(x_f,t|x_0)$ of $X(t)$, i.e., the probability density for the time for the stochastic variable $X(t)$ to take the value $x_f$ for the first time, provided $X(0)=x_0$. This can be accomplished by solving Eq.~\eqref{eq:14} with the initial and final (boundary) conditions \cite{ka07}
\begin{equation}
 \label{eq:15}
 \begin{split}
 &p(x,0)=\delta(x-x_{0}), \\
 &p(x_f,t)=0,
 \end{split}
\end{equation}
and the boundary conditions physically relevant for the problem: absorbing, reflecting, or mixed boundary conditions at $\partial I$, the limit points of $I$. Observe that the absorbing boundary condition at $x_f$, together with the initial condition at $x_0$, allows us to take $x_f\in \partial I$. Put otherwise, in the special case of a one-dimensional problem defined in a given interval $ [a,b] $, where $ a $ and $ b $ are the boundaries, when one considers an absorbing point $ x_f $ the original space reduces to an effective interval $I$, where $I=[a,x_f]$ if $x_0<x_f$ or $I=[x_f,b]$ if $x_0>x_f$.

The solution to the problem~\eqref{eq:14} and \eqref{eq:15} can be formally written as \cite{ga09,dekr95,ri89}
\begin{equation}
 \label{eq:16}
 p(x,t)=\sum_{n=0}^\infty c_n X_n(x)e^{-\lambda_n t}
\end{equation}
where $\{c_n\}_{n=0}^\infty$ are coefficients to be determined by the initial condition, and $\{X_n\}_{n=0}^\infty$ and $\{\lambda_n\}_{n=0}^\infty$ are the associated eigenfunctions and eigenvectors satisfying
\begin{equation}
 \label{eq:17}
 \begin{split}
 &\frac{d}{dx}J[x|X_n]=\lambda_n X_n(x), \\
 & X_n(x_f)=0,
 \end{split}
\end{equation}
with $ J[x\in \partial I|X_n]=0$ if the physical boundaries of the problems are reflecting, $X_n(x\in \partial I)=0$ if they are absorbing, or a combination of both conditions if one is absorbing and the other reflecting.

It is well known from the mathematical theory of the Sturm-Liouville problem \cite{dekr95} that the eigenfunctions form an orthonormal basis of the Hilbert space of square integrable functions defined in $I$, with the scalar or inner product defined as $\medio{f,g}\equiv \int_I dx \, w(x) f(x)g(x)$, and measure $w(x)dx$,
\begin{equation}
 \label{eq:18}
 w(x)=B(x)\exp\left[-2\int^x dx' \frac{A(x')}{B(x')}\right],
\end{equation}
which is proportional to the inverse of the steady-state solution of Eq.~\eqref{eq:14} with reflecting boundary conditions. In Eq.~\eqref{eq:18} it is assumed that the integral is well defined. The scalar product of the Hilbert space, along with the orthonormal properties of $\{X_n\}$, allow us to express the coefficients $c_n$ as a function of the initial condition \eqref{eq:15},
\begin{equation}
 \label{eq:19}
 c_n=\medio{p(x,0),X_n}=w(x_0)X_n(x_0).
\end{equation}

If we now use the general result of Eq.~\eqref{eq:13} for the FP distribution of the problem defined by Eq.~\eqref{eq:14} together with the boundary condition $p(x_f,t)=0$, as explained in the last part of Appendix \ref{app:1}, we get \cite{ka07}
\begin{equation}
 \label{eq:20}
 f(x_f,t|x_0)=\frac{1}{2} B(x_f)\left|\left.\frac{\partial p(x,t)}{\partial x}\right|_{x=x_f}\right|,
\end{equation}
where the absolute value has been introduced in order to unify the notation in the two possible cases $ x_f < x_0 $ and $ x_f > x_0 $, which result in $+$ and $-$ signs, respectively. Using Eq.~\eqref{eq:16}, we obtain
\begin{equation}
 \label{eq:21}
 \begin{split}
 f(x_f,t|x_0)=&\frac12B(x_f) w(x_0)\\ &\times \left| \sum_{n=0}^\infty X_n(x_0)X_n'(x_f)e^{-\lambda_nt}\right|.
 \end{split}
\end{equation}
Note that Eqs.~\eqref{eq:20} and \eqref{eq:21} hold only if $B(x_f)\ne 0$. If the diffusion coefficient vanishes at the final state, we should reconsider the problem \eqref{eq:14} without the boundary condition $p(x_f,t)$ and then use the relation \eqref{eq:13}. 

The eigenvalues of a Sturm-Liouville problem form an ordered sequence, $ 0 \leq \lambda_0 < \lambda_1 < \ldots \, $. Hence, we infer from Eq.~\eqref{eq:21} that the last stage of the dynamics ($t\lambda_s \gtrsim 1$) shows always an exponential decay, with a characteristic time related to the smallest eigenvalue $\lambda_s$ for which $X_s(x_0)X_s'(x_f)\ne 0$. The other limit, namely, $t\lambda_s\lesssim 1$ will be considered in the next section.

\section{General features of the first-passage distribution\label{sec:3}}

Consider the eigenvalue problem of the 1D Fokker-Planck equation \eqref{eq:14}. If we introduce the so-called Liouville-Green transformation \cite{na04}
\begin{equation}
 \label{eq:28a}
 \begin{split}
 &y(x)=\left|\int_{x_f}^x\sqrt{\frac{2}{B(x')}}dx'\right|, \\
 &Y_n(x)=B^{\frac14}(x)w^{\frac12}(x)X_n(x), \quad n=0,1,\dots
 \end{split}
\end{equation}
Eq.~\eqref{eq:17} for the eigenvectors becomes
\begin{equation}
 \label{eq:28b}
 \frac{d^2Y_n(y)}{dy^2}+\left[\lambda_n- \Delta(y)\right] Y_n(y)=0,
\end{equation}
where
\begin{equation}
 \label{eq:28c}
 \Delta(x) \equiv \frac{16(A^2+A'B-AB')+3B'^2-4BB''}{32B},
\end{equation}
with the prime denoting a derivative with respect to $x$. By dimensional analysis, the solution of Eq.~\eqref{eq:28b} can be written as
\begin{equation}
 \label{eq:28c2}
 Y_n(y,\lambda_n,\boldsymbol{\Delta}_c)=g(\sqrt{\lambda_n}y,\boldsymbol{\Delta}_c/\lambda_n).
\end{equation}
where $\boldsymbol{\Delta}_c$ is a constant vector with physical dimensions of $\Delta(x)$ made from all constants $\Delta(x)$ may depend on. The function $g$ must satisfy, according to the boundary conditions and the identity $y(x_f)=0$, the following relations
\begin{equation}
 \label{eq:28d}
 g(0,\boldsymbol{\Delta}_c/\lambda_n)=0,
\end{equation}
and 
\begin{equation}
 \label{eq:28dd}
g\left(\sqrt{\lambda_n}y_*,\boldsymbol{\Delta}_c/\lambda_n \right)=0 \quad \text{or} 
 \quad J\left[\sqrt{\lambda_n}y_*|B^{-\frac14}w^{-\frac12}g\right]=0,
\end{equation}
where
\begin{equation}
 \label{eq:28e}
 y_*\equiv \left \{
 \begin{split}
 y(\max \, I) \quad \text{if} \quad x_0>x_f \\
 y(\min \, I) \quad \text{if} \quad x_0<x_f.
 \end{split}
 \right.
\end{equation}
The solution of \eqref{eq:28d}--\eqref{eq:28dd} determines the set of eigenvalues $\{\lambda_n\}_{n=0}^\infty$ and determines, up to a normalization factor, the eigenfunctions $\{Y_n\}_{n=0}^\infty$ and $\{X_n\}_{n=0}^\infty$.

Recalling that $X(x_f)=0$, which implies $Y(x_f)=0$, and the relations in Eqs. \eqref{eq:28a} and \eqref{eq:28c2}, we have $X_n'(x_f)=2^{\frac12}B^{-\frac34}(x_f)w^{-\frac12}(x_f) \frac{d}{dy}Y_n(x_f)=2^{\frac12}B^{-\frac34}(x_f)w^{-\frac12}(x_f) \sqrt{\lambda_n} \, \overset{\textbf{.}}{g} \left(0,\boldsymbol{\Delta}_c/\lambda_n \right)$ which is proportional to $\sqrt{\lambda_n} \, \overset{\textbf{.}}{g} \left(0,\boldsymbol{\Delta}_c/\lambda_n \right)$, where the overdot denotes a derivative with respect to $y$. With these relations we obtain from Eq. \eqref{eq:21} that
\begin{equation}
 \label{eq:24}
 \begin{split}
 f(x_f,t|x_0)\propto \left| \sum_{n=0}^\infty \right. & \sqrt{\lambda_n}g\left(\sqrt{\lambda_n}y_0,\boldsymbol{\Delta}_c/\lambda_n \right) \\ & \left. \times \overset{\textbf{.}}{g} \left(0,\boldsymbol{\Delta}_c/\lambda_n \right) e^{-\lambda_nt} \vphantom{\sum_{n=0}^\infty} \right|,
 \end{split}
\end{equation}
where
\begin{equation}
 \label{eq:24a}
 y_0\equiv y(x_0),
\end{equation}
which depends also on $x_f$ [see Eq.~\eqref{eq:28a}].

The first-passage distribution Eq.~\eqref{eq:24} can be simplified further if we consider its continuum limit, i.e., write it as an integral on the eigenvalues. We show the details of the derivation and the range of validity of the used approximations in Appendix \ref{app:1_2}. Defining $ s_n = \sqrt{\lambda_n t} $, we obtain that Eq.~\eqref{eq:24} becomes

\begin{equation}
 \begin{split}
 \label{eq:25}
 f(x_f,t|x_0)\sim t^{-1} \left| \int_0^\infty ds \right. & ~g\left(\frac{y_0}{\sqrt{t}}s,\boldsymbol{\Delta}_c\,t/s^2\right) \\ \times & \left. \overset{\textbf{.}}{g}\left(0,\boldsymbol{\Delta}_c\,t/s^2\right)s e^{-s^2} \vphantom{\int_0^\infty} \right|,
 \end{split}
\end{equation}
up to a normalization factor. Observe that the dependence on $y_*$ of Eq.~\eqref{eq:25} has disappeared, meaning that under this approximation we are neglecting the effect of the boundaries of $I$. However, for times $t\gtrsim y_*^2$, the time evolution of the FP distribution does depend on the boundaries and involves only a few terms of the sum of \eqref{eq:24}.

At this point, we have obtained a general expression for the FP distribution from point $x_0$ to $x_f$, valid for bounded $\Delta$, which approximates well the actual distribution if we are not in the latest stages of the dynamics. It will be used next to infer more explicit expressions.

\subsection*{Case of bounded $\Delta$}

If $\Delta(x)$, or equivalently $\Delta(y)$, is a bounded function for the allowed values of $x$, we can disregard its contribution in the eigenvalue problem of Eq.~\eqref{eq:28b} as a first approximation. This is valid for $\lambda>\Delta_*$. This is the so-called WKB approximation \cite{na04} and gives
\begin{equation}
 \label{eq:25a}
 \begin{split}
 & Y_n(y) \propto \sin\left(\sqrt{\lambda_n}y\right), \\
 & \lambda_n =\frac{n^2\pi^2}{y_*^2},
 \end{split}
\end{equation}
for $n$ large enough. The smaller the value of $\Delta_*$ is, the better the WKB approximation works for the FP distribution, as shown below. The relation $g(\sqrt{\lambda_n}y,\boldsymbol{\Delta}_c/\lambda_n) \simeq g(\sqrt{\lambda_n}y,0)$ is a good approximation to Eq.~\eqref{eq:24} if the contribution of the smallest eigenvalues (modes) can be neglected for the times of interest. More precisely, by looking at Eq.~\eqref{eq:25} we realize that the WKB approximation is equivalent to the limit $\boldsymbol{\Delta}_c\,t/s^2\to 0$. However, for a bounded $\Delta(x)$, $g$, and $\overset{\textbf{.}}{g}$ the important contributions to Eq.~\eqref{eq:25} occur around $s = 1/\sqrt{2}$, the location of the maximum of $se^{-s^2}$, that is, the range of validity of the WKB reduces to $\Delta_* t\ll 1$, with $\Delta_* \equiv \max_{x}|\Delta(x)|$, and as an estimation we take $t<t_{\,\text{WKB}}\equiv \Delta_*^{-1}$. This condition is consistent with the continuum limit required to obtain Eq.~\eqref{eq:25}.

Now, from Eqs.~\eqref{eq:25} and \eqref{eq:25a} under the WKB approximation ($t<t_{\,\text{WKB}}$), the first-passage distribution reduces to the L\'evy-Smirnov density \cite{gome16}
\begin{equation}
 \label{eq:27}
 f(x_f,t|x_0)\simeq \frac{y_0}{2\sqrt{\pi}} t^{-3/2}e^{-\frac{y_0^2}{4t}}.
\end{equation}
Note that $y_0$ depends not only on $x_0$, but also on $x_f$ [see Eqs.~\eqref{eq:28a} and \eqref{eq:24a}]. Equation \eqref{eq:27} indicates that the FP distribution vanishes exponentially fast near $t=0$ and more interestingly shows a power-law decay with exponent $-3/2$ for larger times, regardless of the spatial dependence of the drift and diffusion terms. The range of times where the power law is obtained is
\begin{equation}
 \label{eq:27a}
 y^2_0\lesssim t \lesssim y^2_*,
\end{equation}
which depends only on the diffusion coefficient. As already discussed, Eq.~\eqref{eq:27} breaks down for $t>t_{\,\text{WKB}}$, where a few modes, the lowest ones, dominate the dynamics. The decay is exponential for $t\lambda_0>1$, an inequality that in this case depends on both the diffusion and drift terms. Moreover, a nonuniversal scaling can also emerge just before the exponential decay, as can be realized if we were to include the $\Delta$-dependent corrections of the WKB approximation. See Refs. \cite{gome16} and \cite{sc10} for a deeper and more general study of the exponential and preexponential decay.

\begin{figure}[!t]
 \centering
 \includegraphics[width=\linewidth]{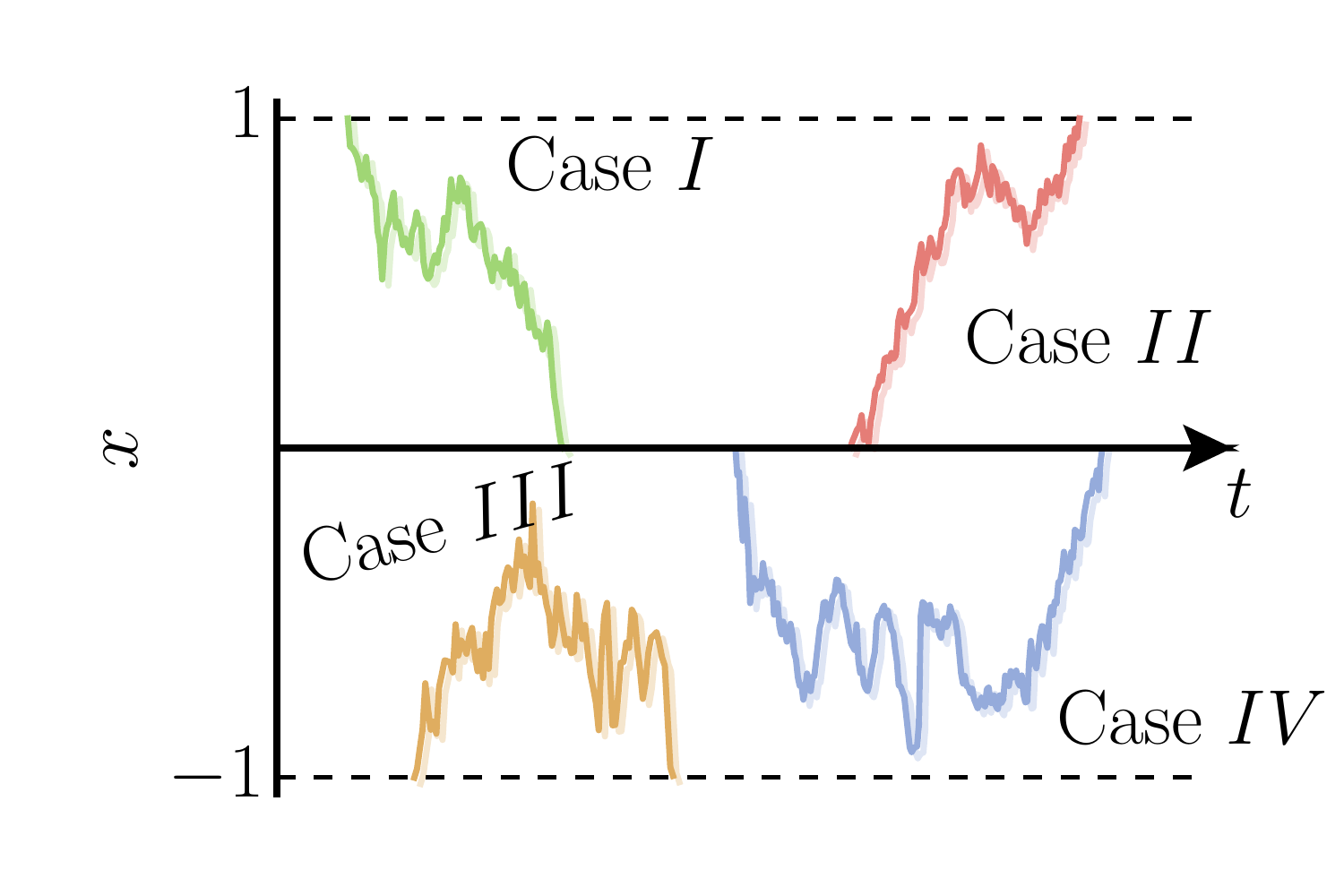}
 \caption{Schematic representation of the types of trajectories we study along this work. Cases I and II correspond to the FP from the boundary to the center and from the center to the boundary, respectively. Cases III and IV correspond to the FR to the boundary and to the center, respectively. In practice, however, the latter cases are considered as FP from $x_0=1-1/N$ to $x_f=1$ and from $x_0=1/N$ to $x_f=0$, respectively.}
 \label{fig:2a}
\end{figure}

\begin{figure*}[!t]
 \centering
 \includegraphics[width=0.24\linewidth]{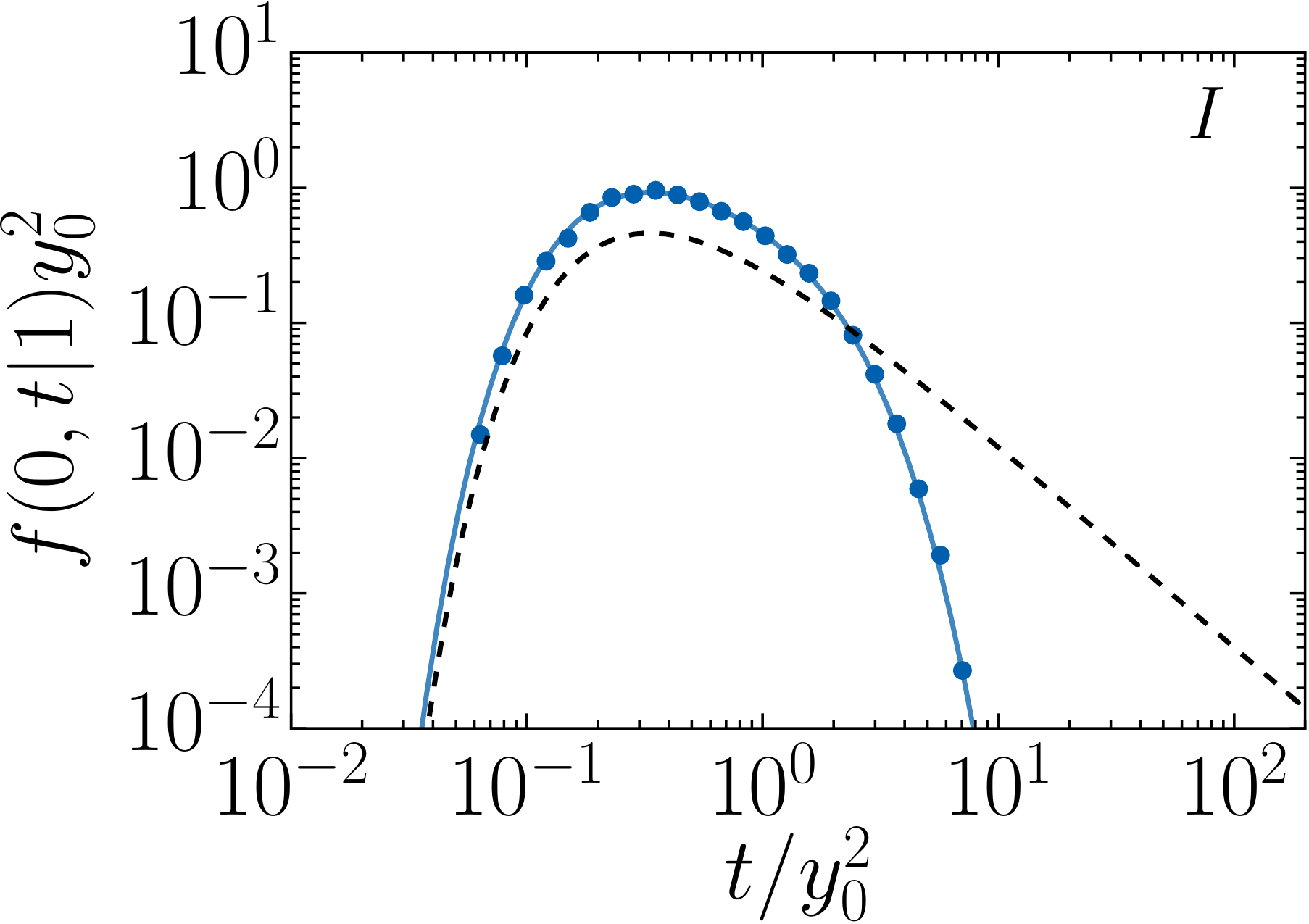}
 \includegraphics[width=0.24\linewidth]{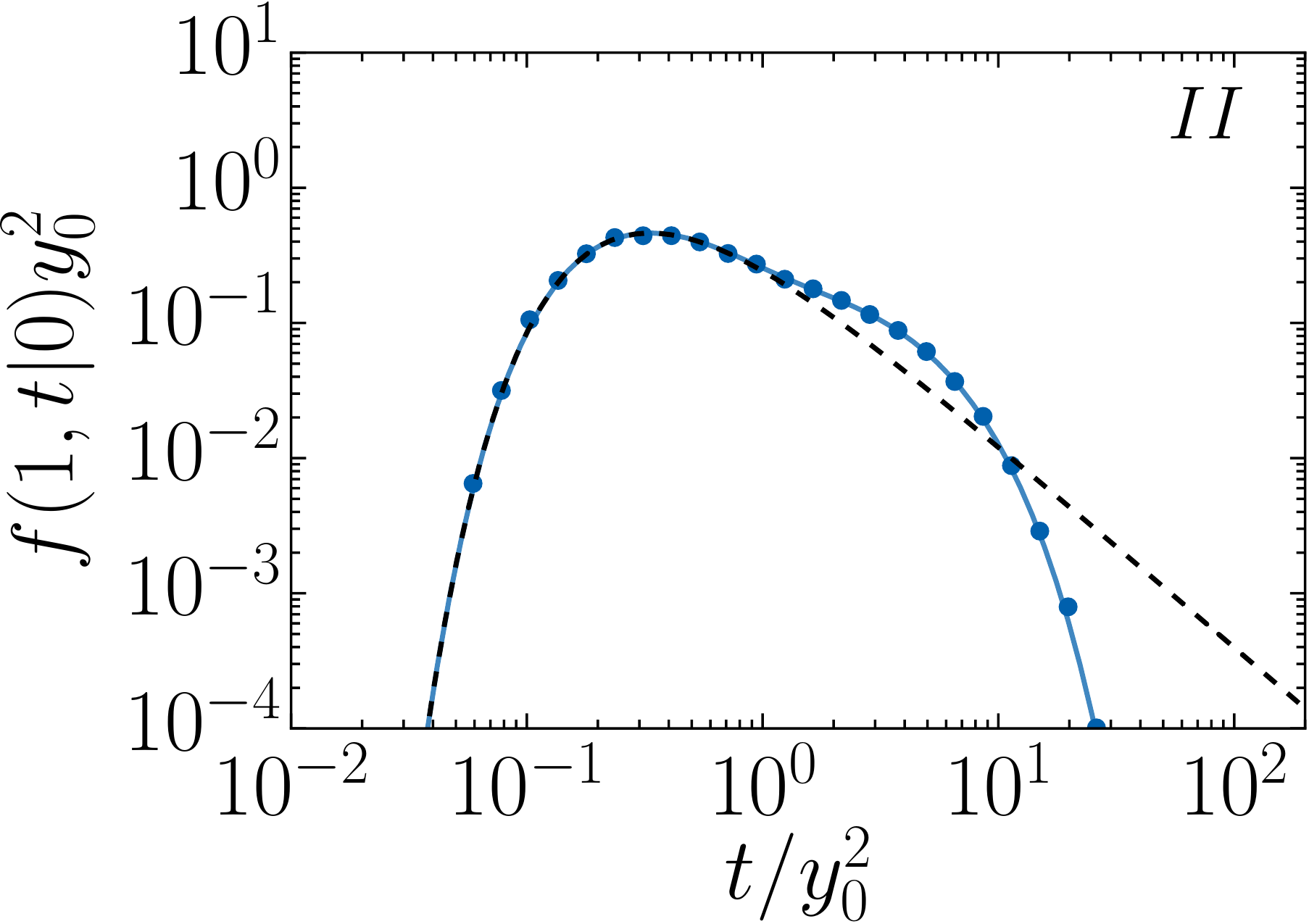}
 \includegraphics[width=0.24\linewidth]{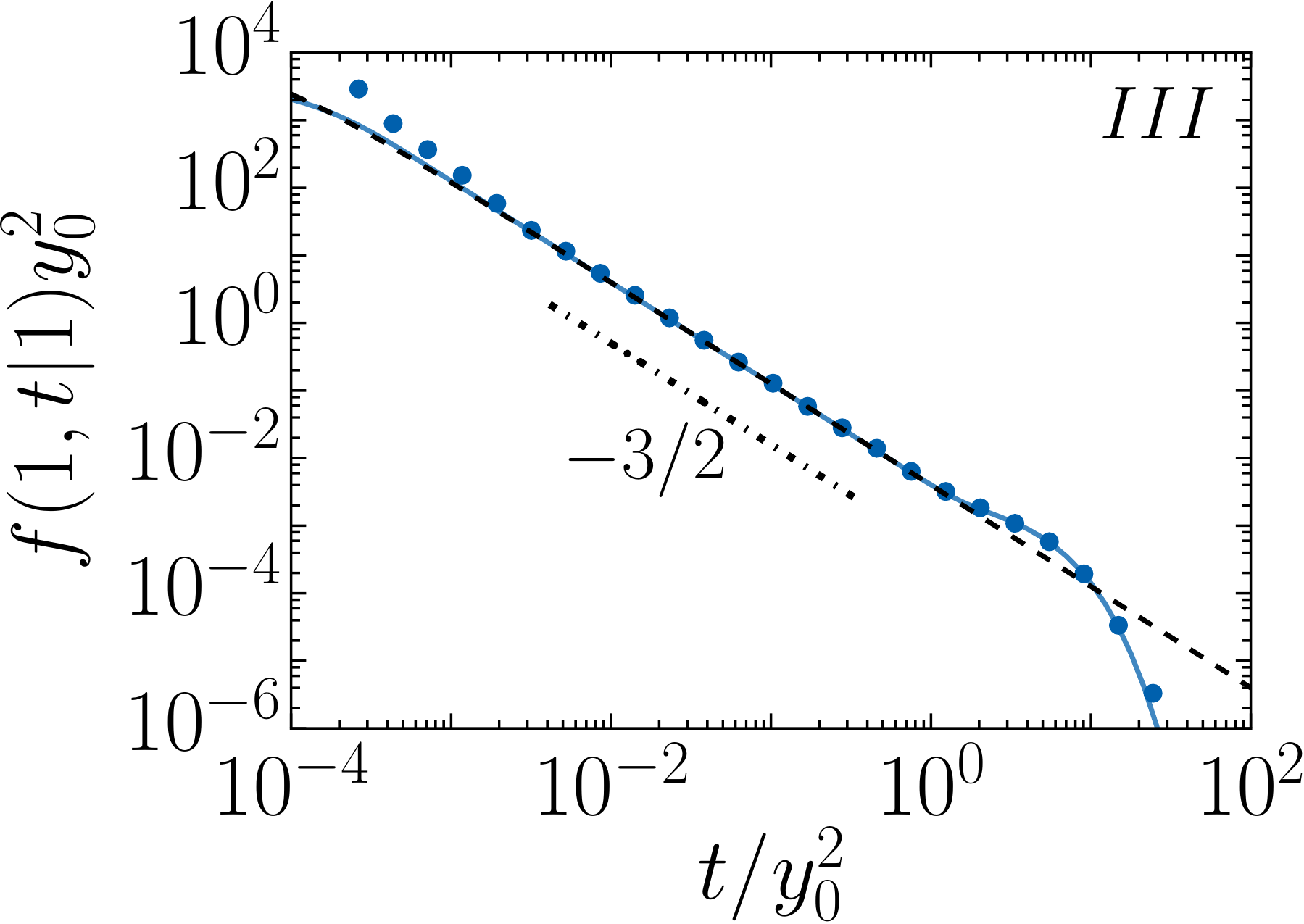}
 \includegraphics[width=0.24\linewidth]{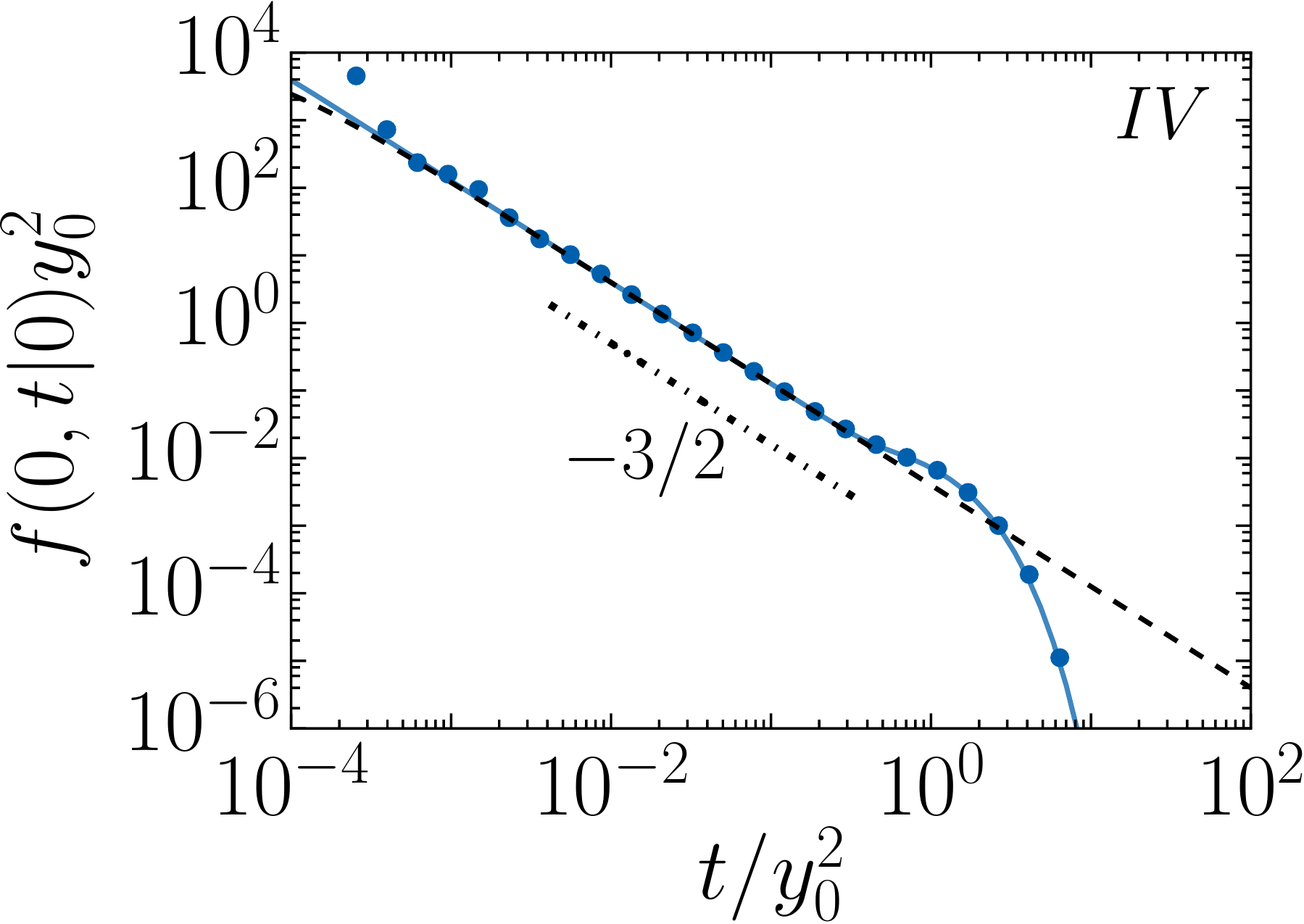}
 \caption{First-passage and first-return distributions for a random walker in the interval $[-1,1]$, with $ N = 200 $ subdivisions. From left to right, the cases I to IV shown in Fig. \ref{fig:2a}. Symbols are from Monte Carlo simulations, solid lines are from the solutions of Eqs.~\eqref{eq:a3_fpt} and \eqref{eq:fret} (using $ 10^5 $ addends) and dashed lines are the approximate theory of Eq.~\eqref{eq:27}.}
 \label{fig:2}
\end{figure*}

Our first example of the class of models with a bounded $ \Delta(x) $ is the random walk within a finite interval whose continuous version is the Brownian motion. In the terminology of the Fokker-Planck equation [see Eq.~\eqref{eq:14}], it is described by a null drift term $A=0$ and a constant diffusion $B$ \cite{sato00}. The discrete random walk, the one we simulate, is defined as follows. Let $ x $ be the position of the walker, where $ x \in [-1,1] $. At each time step $ \delta t $ the walker moves with equal probability to adjacent states $ x \to x \pm \delta x $. The step size is $ \delta x = 2/N $, meaning that the $[-1,1]$ interval is divided into $ N $ subunits. The time is measured in Monte Carlo steps (MCS), i.e., $ N $ jumps correspond to 1 MCS. This leads to $ \delta t = 1/N $. The equivalence between the continuous and the discrete versions is achieved when $ N \gg 1 $. Performing the expansion for large $ N $ in the master equation of the random walk, one readily obtains the diffusion coefficient $B=\delta x^{2}/\delta t=4/N $.

We study the FP and FR time distributions in several situations, as sketched in Figure \ref{fig:2a}. Cases I and II correspond to FP processes. Case I corresponds to trajectories from the extreme of the interval to its center. Case II takes into account trajectories from the center of the interval to one of the boundaries. Cases III and IV are scenarios of FR to the boundary and to the center, respectively. If the drift and the diffusion terms are symmetric with respect to the center, the distributions are independent of which boundary we depart from or we arrive at.

In the context of random walks, the distributions for the four cases can be obtained analytically. For the sake of completeness we summarize this calculation in Appendix \ref{app:2}. In Fig. \ref{fig:2} we display the exact distributions, together with their WKB approximation of Eq.~\eqref{eq:27} and the Monte Carlo simulations of the discrete random walker (RW) process. In the region of small and intermediate times, we find almost perfect agreement between simulations and the approximation. This was expected since in this case the WKB approximation provides the exact eigenvalues and eigenvectors (for absorbing boundary conditions). The differences at small times are due to the difference between the Monte Carlo dynamics and that of the Fokker-Planck equation (since $t\sim \delta t$). In the long-time limit $t\lambda_0>1$, the continuum approximation fails, as already analyzed. Specifically, it does not capture the latest exponential decay, since the smallest eigenvalues have been disregarded. Observe that there are no power laws for the first two plots, since for the corresponding values of the parameters we have $y_0 \simeq y_*$ and, according to Eq.~\eqref{eq:27a}, the power-law region disappears.

\begin{figure*}[!t]
 \centering
 \includegraphics[width=0.24\linewidth]{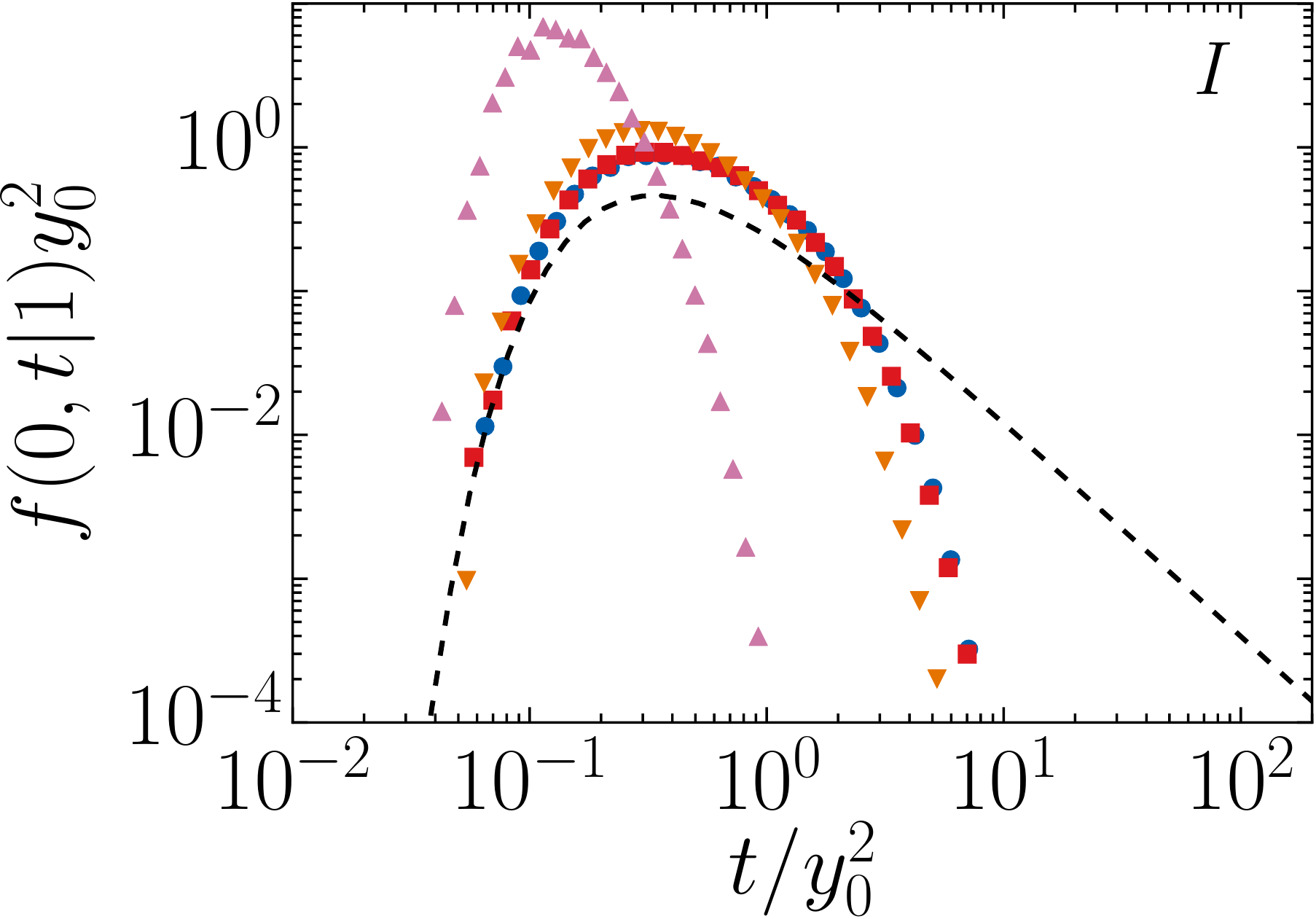}
 \includegraphics[width=0.24\linewidth]{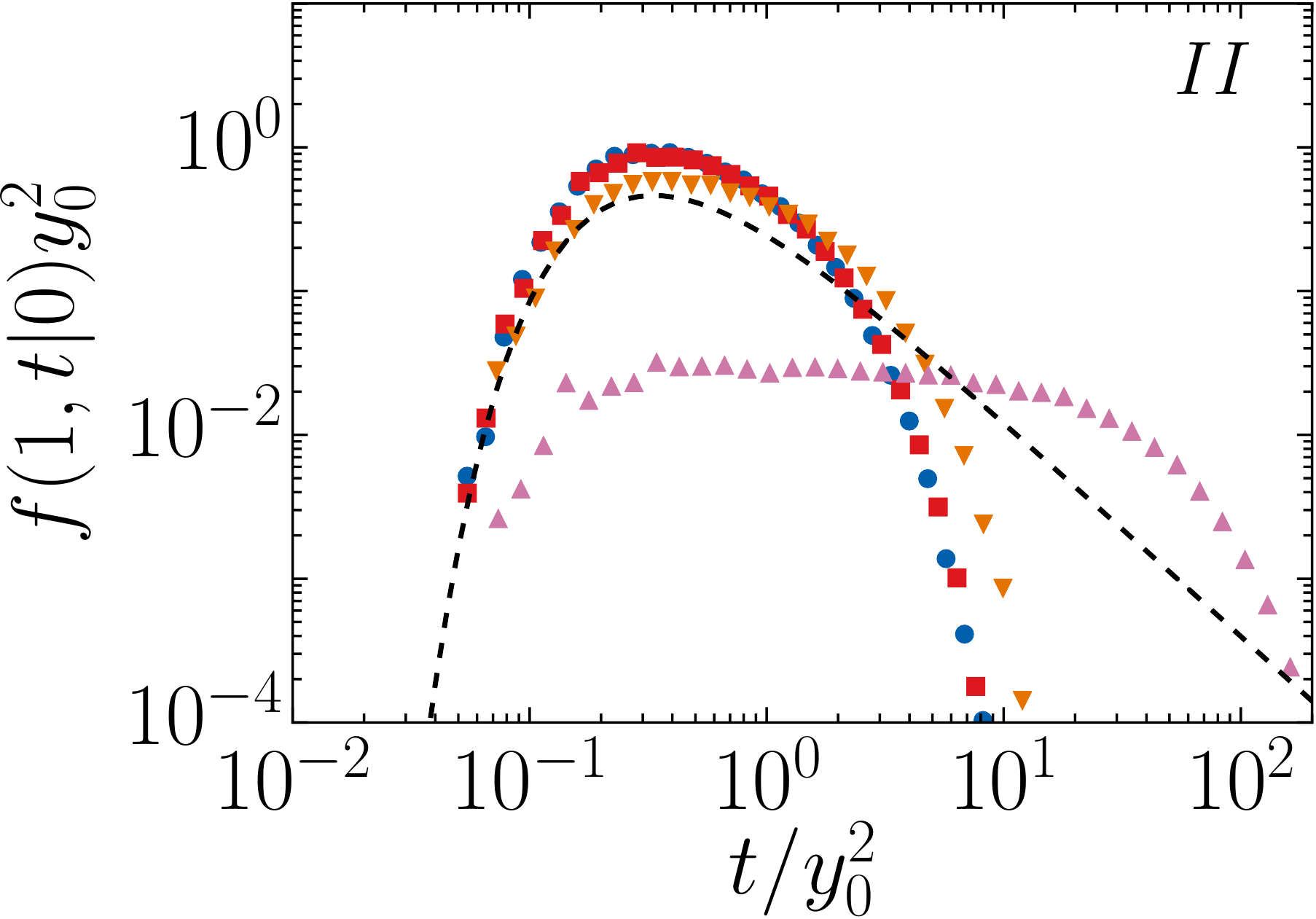}
 \includegraphics[width=0.24\linewidth]{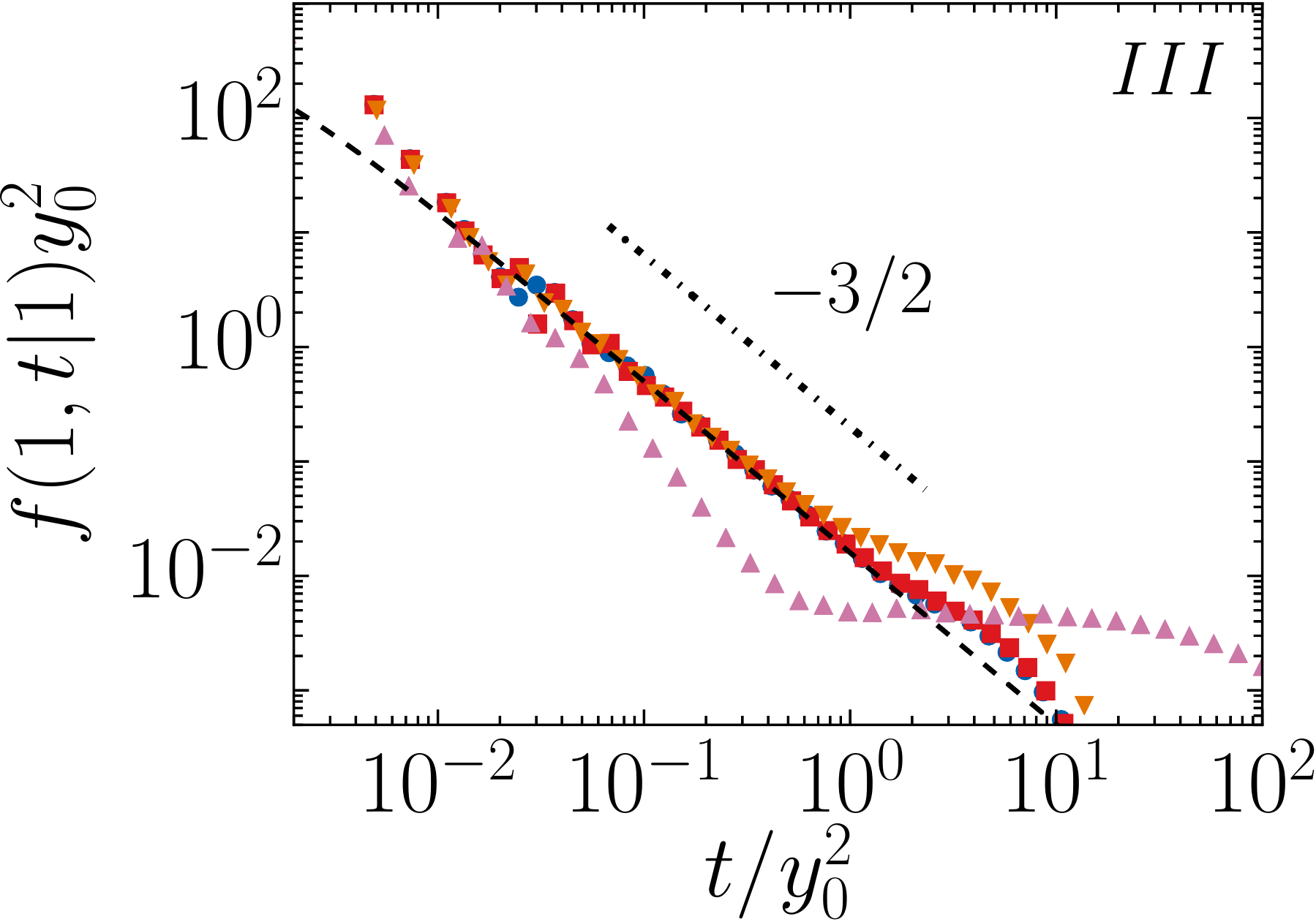}
 \includegraphics[width=0.24\linewidth]{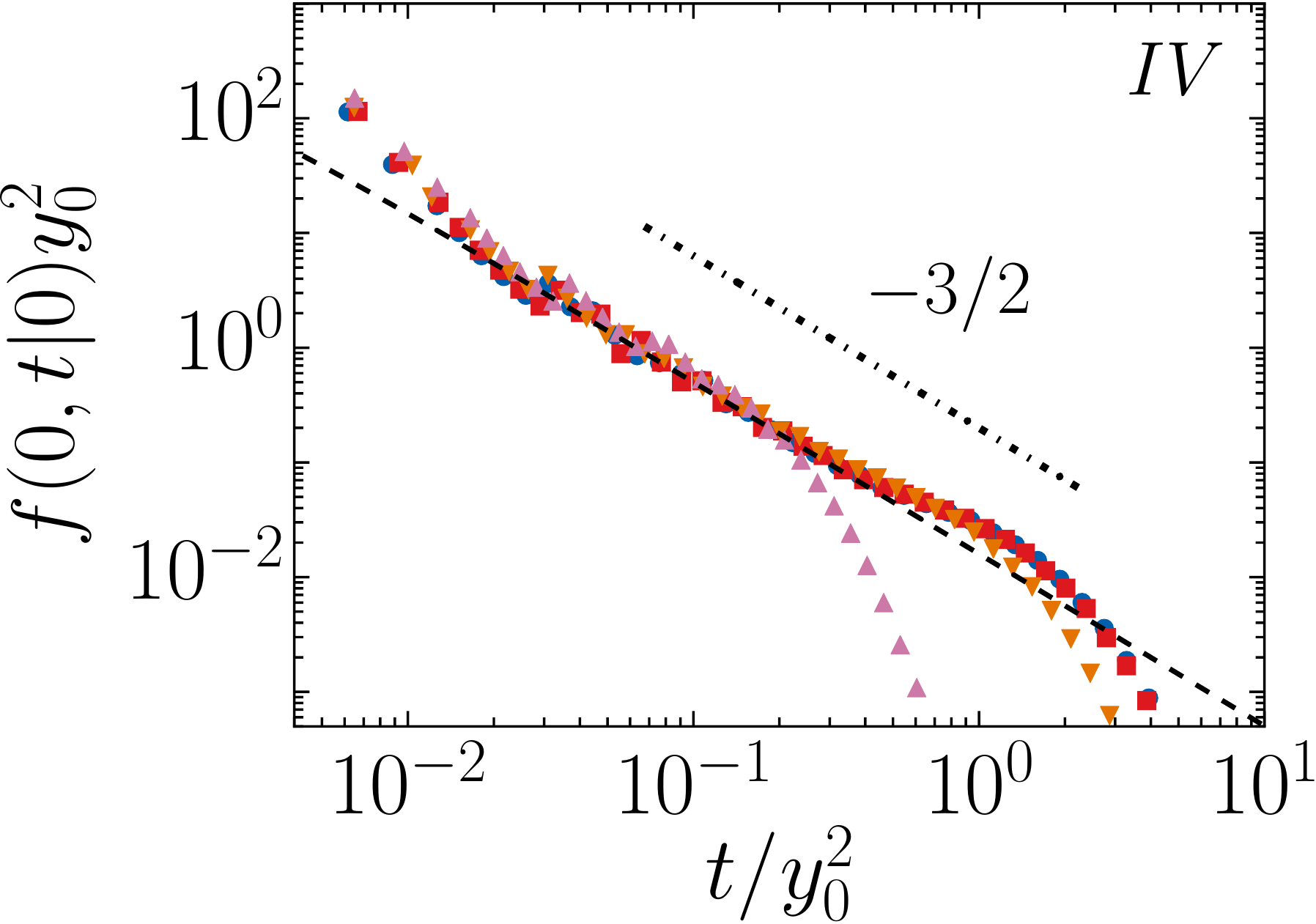}
 \caption{First-passage and first-return distributions for the Ornstein-Uhlenbeck process, Eq.~\eqref{eq:b1}, in the interval $[-1,1]$, with $ N = 50 $ subdivisions. The four cases are those of Fig. \ref{fig:2a}. Symbols correspond to Monte Carlo simulations: $k=6.25$ (pink up triangle), $1.25$ (orange down triangle), $0.125$ (red square), $0.0125$ (blue circle). The dashed line is the approximate theory of Eq.~\eqref{eq:27}.}
 \label{fig:2b}
\end{figure*}

Another interesting example is the Ornstein-Uhlenbeck process (see chapter 8 of \cite{makalu09}). It is characterized by a constant diffusion coefficient and a linear position-dependent drift, so the function $ \Delta(x) $ never diverges in a finite interval. For $x\in[-1,1]$, the drift and diffusion coefficients are written as
\begin{equation}
 \label{eq:b1}
 \begin{split}
 & A(x)=-\frac{k}{2}x, \\
 & B(x)=\frac{1}{N},
 \end{split}
\end{equation}
where $k$ is a constant and $N$ is the number of subunits of the interval in the Monte Carlo simulations. Note that for $k>0$ and $x(0)=0$ this process describes the decay from a linearly unstable state (case II). The main results are summarized in Fig. \ref{fig:2b} where the same four cases of Fig. \ref{fig:2a} are considered. First-passage distributions have a well-defined peak, whose position depends on the strength of the drift $k$. Thus, the transitions from the boundary to the center [Fig. \ref{fig:2b}I] for high $k$ are faster than those for small $k$. The opposite is found for transitions from the center to the boundary [Fig. \ref{fig:2b}II]. These results are reflected in the shape of the distributions. Regarding the FR distributions [Figs. \ref{fig:2b}III and \ref{fig:2b}IV] we see that although the Ornstein-Uhlenbeck process has a drift, the power laws are still present, with the same exponent as that of the random walk. The approximate theory, given by Eq.~\eqref{eq:27}, compares well for all values of the diffusion coefficient, except for the two extreme values of $k$, since $t_{\,\text{WKB}}\sim y_*^2/10$. In these latter cases, the approach to be considered in the next subsection is needed.

\subsection*{Case of unbounded $\Delta$}
The results of the preceding subsection require $\Delta(x)$ to be bounded. If this is not the case, i.e., if the drift term is not bounded and/or the diffusion coefficient vanishes at or near the boundaries of the interval, the FP distribution may still exhibit a power-law behavior but with an exponent different from $-3/2$ (see \cite{br00} and \cite{co08}). This is the case of neuron avalanches \cite{mahimasaplmu17} and other significant models in nonequilibrium statistical mechanics \cite{savibumu17}, in which the first-passage distribution from a point nearby the absorbing state toward the absorbing state itself follows a power law of exponent $-2$. Here we show that the result holds for any drift and diffusion coefficients such that $\Delta(x)$, defined in \eqref{eq:28c}, can be written as
\begin{equation}
 \label{eq:34}
 \Delta(x)=\Delta_r(x)+\frac{\Delta_s}{x-x_s},
\end{equation}
where $\Delta_r$ is a bounded function, the singularity $x_s$ coincides with one of the borders of $I$ or $x_s\notin I$ but close to $I$, and $\Delta_s$ is a constant that can be positive or negative. The interval $I$ where the system effectively evolves was defined just after Eq.~\eqref{eq:15}.

Following similar arguments behind the WKB approximation for the bounded case, we can solve the eigenvalue problem with $\Delta(x)$ given by Eq. \eqref{eq:34} by properly approximating the function $\Delta$. Since $\Delta_r(x)$ is a bounded function, for large enough $\lambda$ the only relevant part of $\Delta(x)$ for our purposes is the one with the singularity, $\Delta(x)\sim \Delta_s/(x-x_s)$. This implies that different functions $\Delta_r(x)$ provide the same scaling exponents of the FP and FR distributions for $\lambda\gg \Delta_*^{r}\equiv \max_{x}|\Delta_r(x)|$. This fact allows us to select a convenient, analytically tractable bounded function $\Delta_r(x)$. Note that this conclusion justifies the generality of the results to be derived in the following. Although the most natural procedure would be to drop $\Delta_r(x)$, it turns out that the resulting problem is very hard to tackle analytically. Instead, one interesting option is to consider $A(x)=0$ and $B(x)=(1-x^2)/(2N)$, with $N$ a constant and $x\in [-1,1]$. These coefficients correspond to those of the voter model, to be explained in detail later on. The diffusion coefficient vanishes at the boundaries of the interval, i.e. we have two absorbing states. Due to symmetry properties, however, we can reduce the problem to $x \in [0,1] $ and study the effects of only one absorbing state, namely $ x = 1 $. Now we have $\Delta(x)=\frac{x^2+2}{16N(1-x^2)}=\frac{1-2x}{32N(1+x)}+\frac{3}{32N(1-x)}$, which has been separated in the form  of Eq. \eqref{eq:34}. Hence $\Delta_*^{r} = \frac{1}{32N}$, and $\Delta_s=\frac{3}{32N}$. The eigenvalue problem of the voter model can be analytically solved \cite{ki64,slla03,mcwa07}, with the result
\begin{align}
 \label{eq:35}
 &X_n(x)\propto N^{1/4}\sqrt{\frac{n+3/2}{(n+1)(n+2)}} C_n^{3/2}(x)\simeq \frac{C_n^{3/2}(x)}{(2\lambda_n)^{1/4}}, \nonumber \\
 &\lambda_n=\frac{(n+1)(n+2)}{4N} \simeq \frac{n^2}{4N},
\end{align}
where $C_n^{3/2}$ is the Gegenbauer polynomial of order $3/2$ and degree $n$. The approximate relations hold when $n\gg 1$.

\begin{figure*}[!t]
 \centering
 \includegraphics[width=0.24\linewidth]{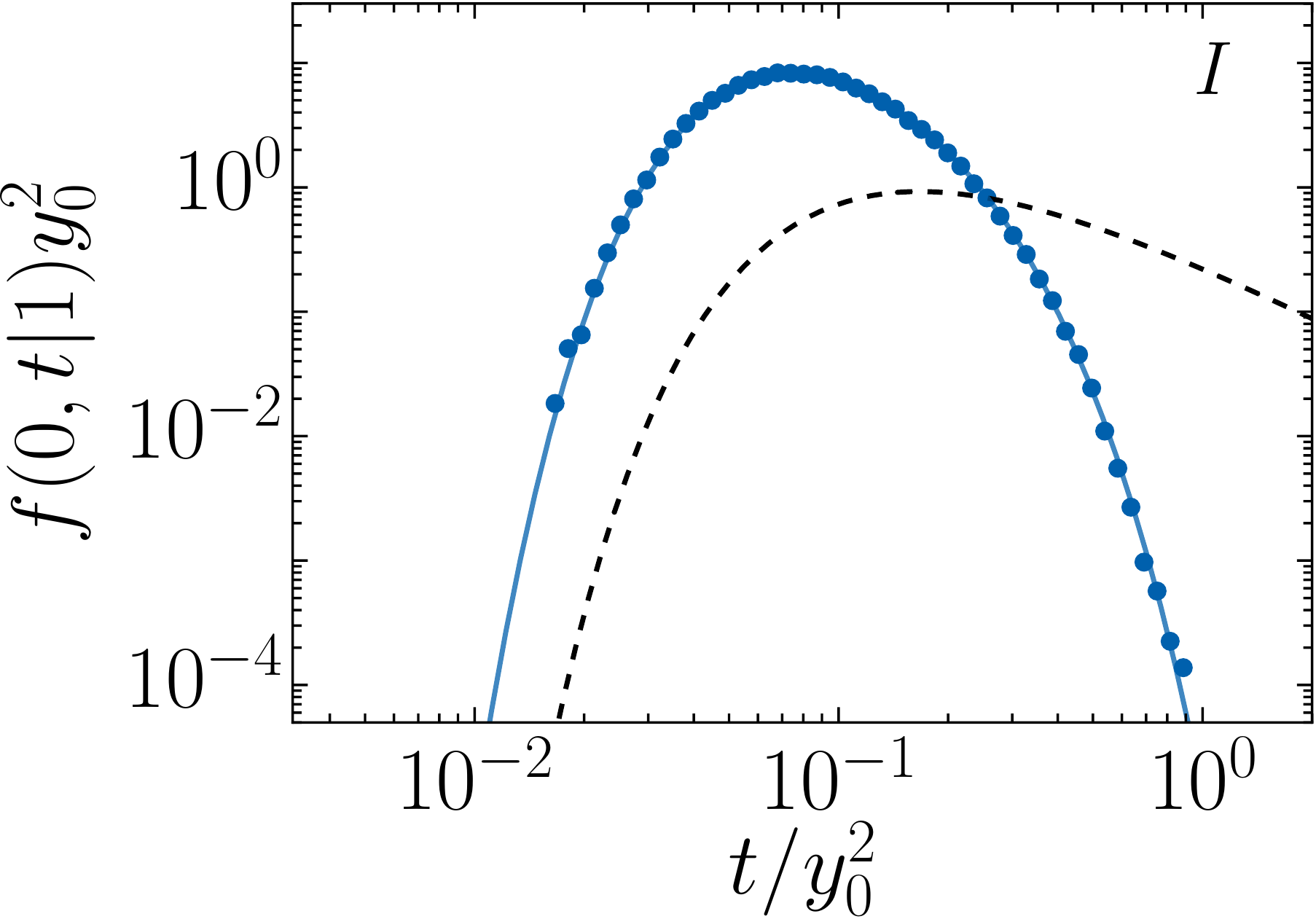}
 \includegraphics[width=0.24\linewidth]{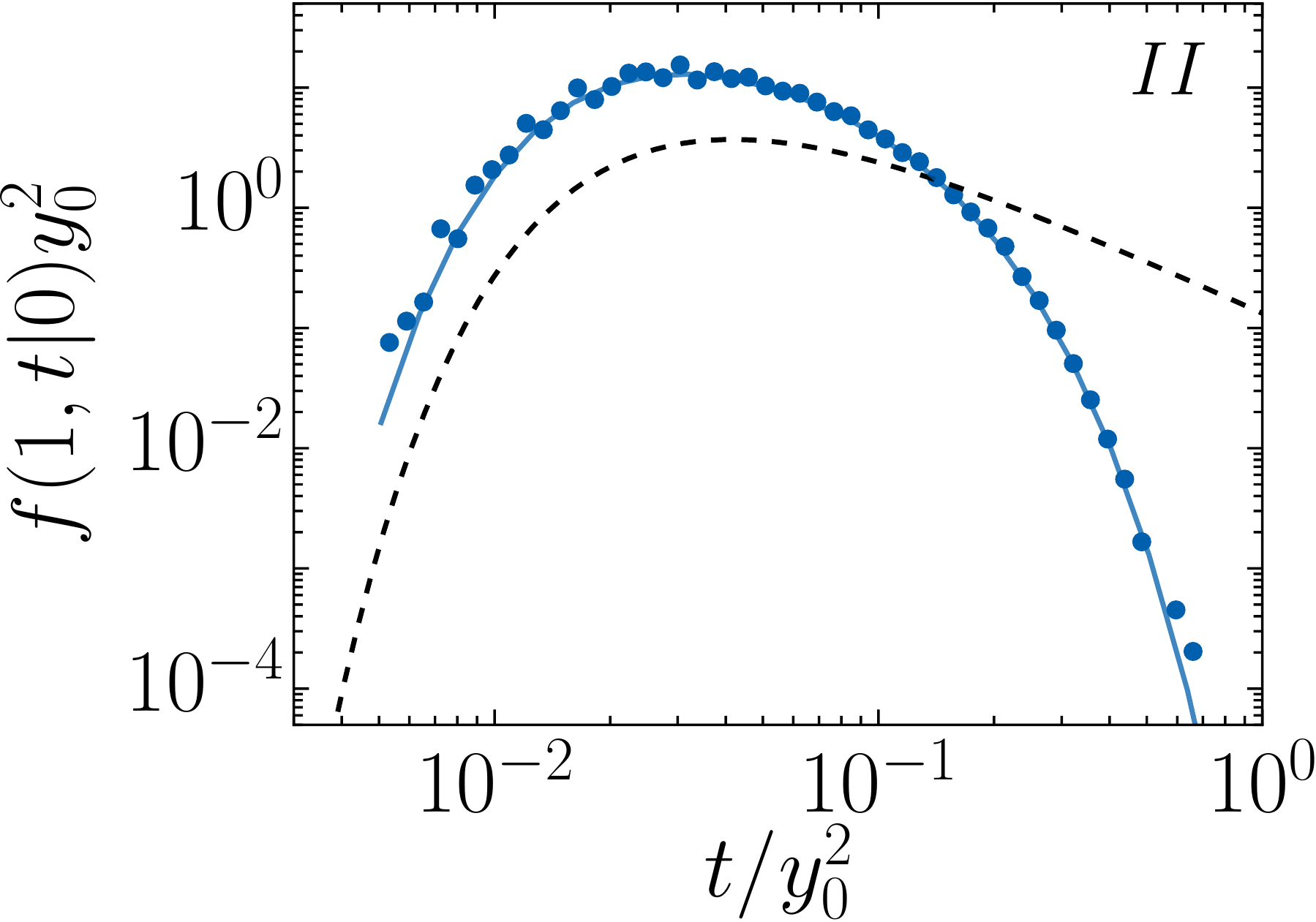}
 \includegraphics[width=0.24\linewidth]{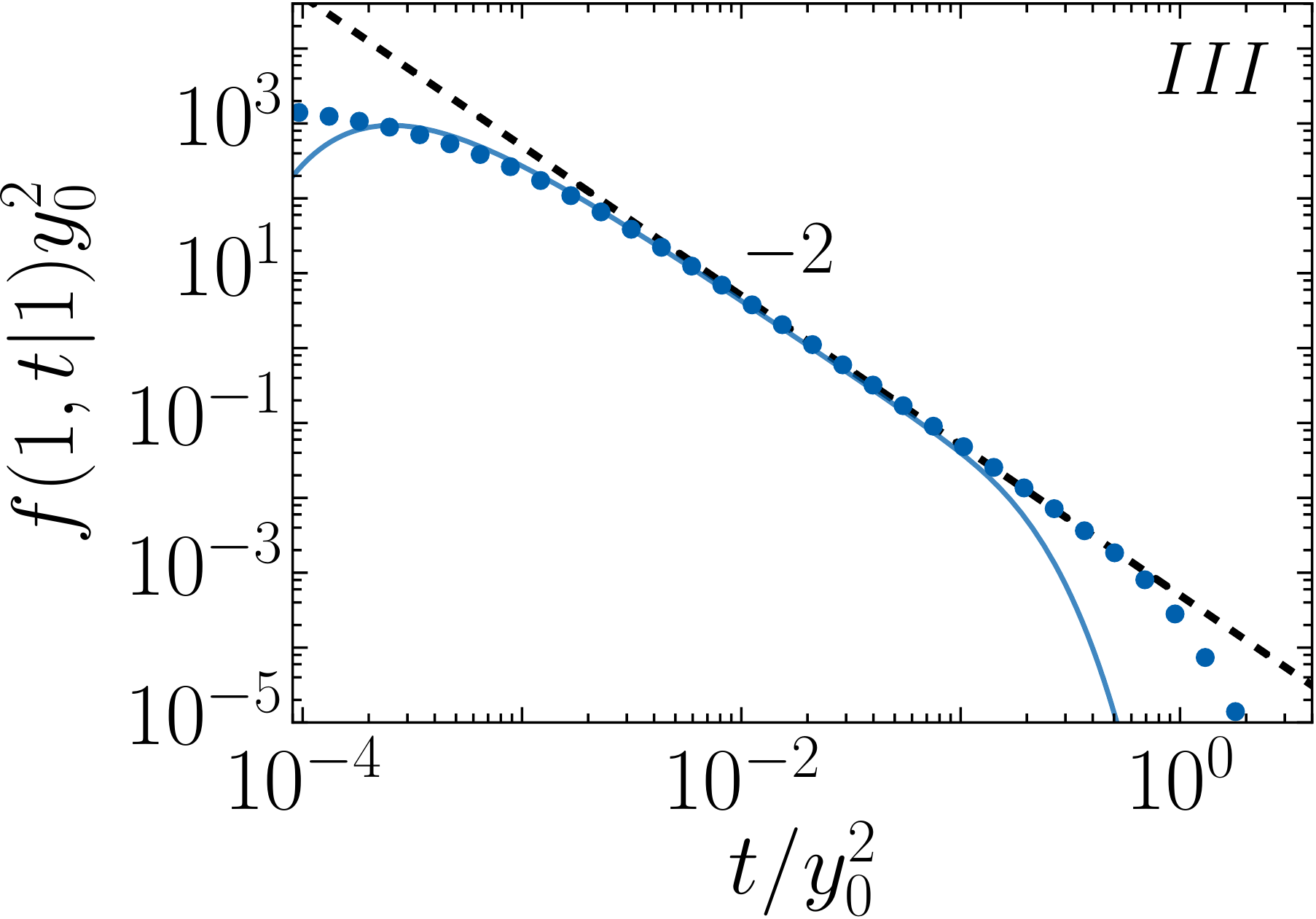}
 \includegraphics[width=0.24\linewidth]{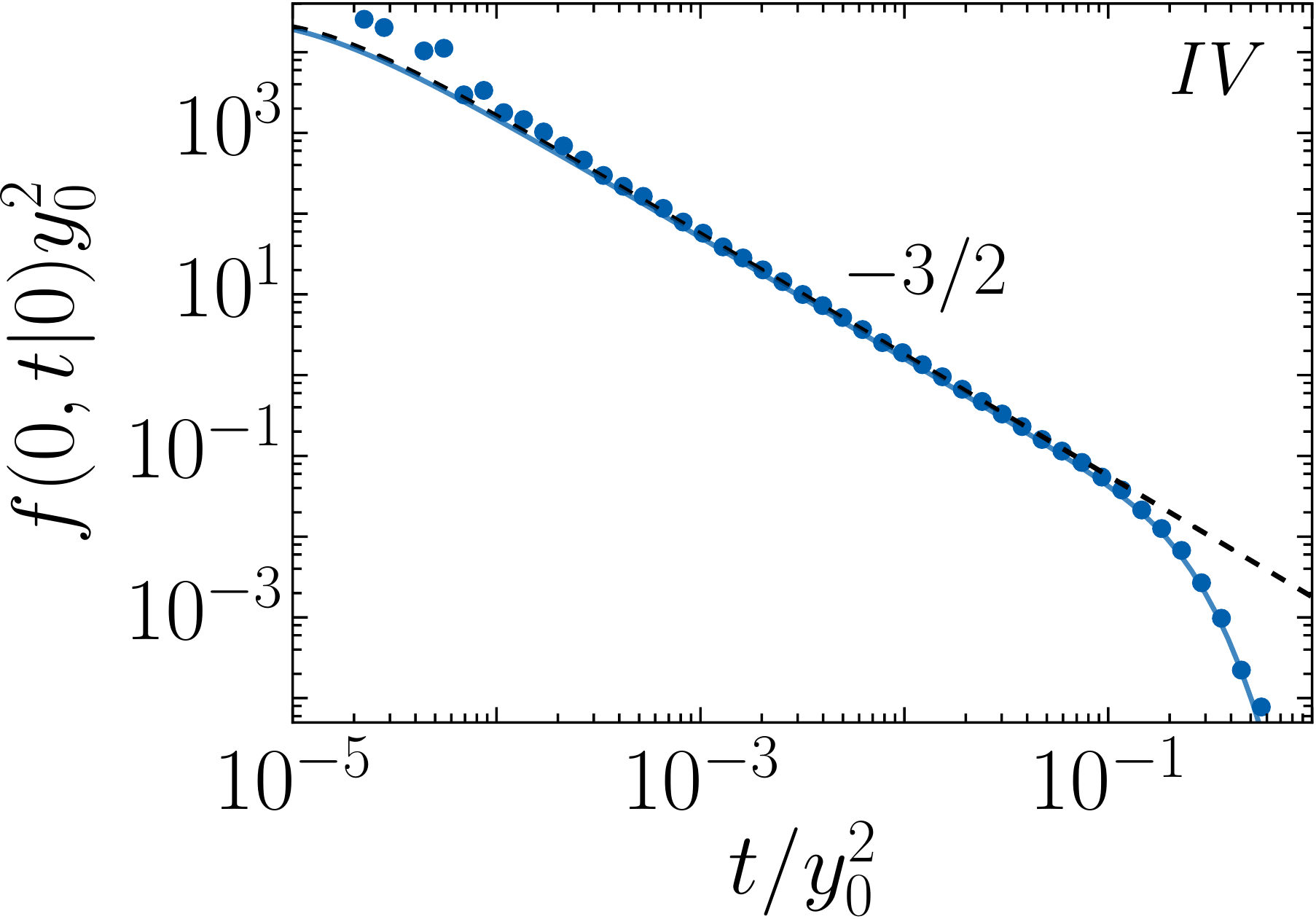}
 \caption{First-passage and first-return distributions studied in the main text for the voter model. Figures correspond to the cases sketched in Fig. \ref{fig:2a}. Symbols are from Monte Carlo simulations for $N=100$ (first two plots) and $N=200$ (last two plots). Solid lines are from the exact theory of Eq.~\eqref{eq:21} (200 addends) and dashed lines are the approximate theory of Eq.~\eqref{eq:27}.}
 \label{fig:3}
\end{figure*}

The results of Eqs.~\eqref{eq:35} can be used now to compute the FP distribution of the four cases shown in Fig. \ref{fig:2a}.
\begin{itemize}
\item Case I $\left( 0=x_f<x_0\sim 1 \right)$: the initial state is close to the singularity and the final state is away from it. Using $C^{3/2}_n(1)=\frac{\Gamma(n+3)}{2\Gamma(n+1)}=2N\lambda_n$ with Eq.~\eqref{eq:36}, we obtain
\begin{equation}
 \label{eq:38}
 f(0,t|1^-)\sim \sum_{n} n^{5/2}\sin\left(\frac{n\pi}{2}\right) e^{-\frac{n^2}{4N} t}\sim t^{0},
\end{equation}
where $t^0$ indicates no power-law decay.
\item Case II $\left( 0=x_0<x_f\sim 1 \right)$: the initial state is far from the singularity while the final state is close to it. Since the final state is now near or at the singularity, we cannot use the result of Eq.~\eqref{eq:21}, but instead we should compute the general relation \eqref{eq:13} without $p(x_f,t)=0$. Since $B(x=1)=0$, now $J[1^-|X_n]=-1/2B'(1)X_n(1^-)$, and
\begin{equation}
 \label{eq:39}
 f(x_f,t|x_0)\sim \sum_n \lambda_n^{-\frac{1}{2}}C_n^{3/2}(x_0)C_{n}^{3/2}(1^-)e^{-\lambda_n t}.
\end{equation}
Proceeding as in case I, we arrive at the same functional dependence of Eq.~\eqref{eq:38}.
\item Case III $\left( 1\sim x_0<x_f=1 \right)$: the initial and final states are close to the singularity. Particularizing Eq.~\eqref{eq:39} for $x_0=1$, we have
\begin{equation}
 \label{eq:40}
 f(x_f,t|x_0)\sim \sum_{n} n^{3} e^{-\frac{n^2}{4N} t}\sim t^{-2},
\end{equation}
for $x_0-x_f\lesssim t\lesssim 1$. This is the power law with exponent $-2$ observed in Ref. \cite{mahimasaplmu17}, which we now realize appears for any drift and diffusion terms of the class of models defined by Eq.~\eqref{eq:34}.
\item Case IV $ \left( 0=x_f<x_0\ll 1 \right)$: the final and initial states are far from the singularity ($x=1$). From Eq.~\eqref{eq:21}, the FP distribution is
 \begin{equation}
 \label{eq:36}
 f(0,t|x_0)\sim \sum_n \lambda_n^{-\frac{1}{2}}C_n^{3/2}(x_0)C_{n-1}^{5/2}(0)e^{-\lambda_n t},
 \end{equation}
where we have used the relation $\frac{d}{dx}C_n^{3/2}(x)=3C_{n-1}^{5/2}(x)$. If $t$ is larger than the typical time the system needs to relax from $x_0$ to $x_f$, namely if $t/N>x_0$, with good approximation we can replace $C_n^{3/2}(x_0)$ by $C_n^{3/2}(x_0)\simeq 3C_{n-1}^{5/2}(0)x_0$. Now, by using $|C_{n-1}^{5/2}(0)|=\frac{4}{3\sqrt{\pi}}\frac{\Gamma\left(\frac{n+4}{2}\right)}{\Gamma\left(\frac{n+1}{2}\right)}\delta_{n,odd}\sim n^{3/2}\delta_{n,odd}$, we obtain
 \begin{equation}
 \label{eq:37}
 f(0,t|x_0)\sim \sum_{n\text{ odd}} n^2 e^{-\frac{n^2}{4N} t}\sim t^{-3/2},
 \end{equation}
where the last approximation holds for $x_0\lesssim t/N\lesssim 1$. Hence, in this scenario, we recover the results of the WKB approximation for the case of bounded $\Delta(x)$.
\end{itemize}
Table \ref{tab:2} summarizes the predictions for the FP distributions obtained so far.

\begin{table}[!b]
 \includegraphics[width=0.8\linewidth]{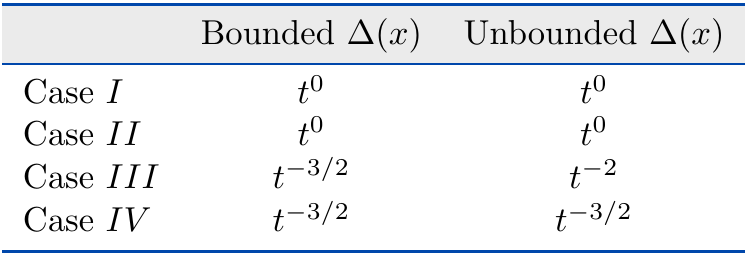}
 \caption{Summary of the theoretical predictions for the asymptotic behavior of the FP distribution for small and intermediate times. The cases are related to those sketched in Fig. \ref{fig:2a}.}
 \label{tab:2}
\end{table}

For other initial and final values of $x$, the situation can be more complicated. For instance, if $x_0\sim x_f$ and close, but not too close (as in case III), to a singularity, the first-passage (-return) distribution can exhibit two power-law decays: the first one at early times with exponent $-3/2$ that accounts for an initial exploration far from the singularity and a later one with exponent $-2$ coming from contributions of the region close to the singularity. The sequence of power laws seems counter intuitive, but it is due to the fact that the system needs more time to leave the singular region. In order for the last stage to appear, the typical time for the system to leave the neighborhood of the singularity should be smaller than $\lambda_0^{-1}$, the smallest time scale of the system. We will illustrate this double-power-law behavior, as well as the predictions of Eqs.~\eqref{eq:38}--\eqref{eq:37} in the next section, after introducing a voterlike family of models. We should mention that the existence of a double scaling is not new and has been reported previously, for example, in the family of Bessel processes \cite{gome16} or in the birth-death process with population-proportional rates \cite{foprmode15}. The latter case has a singularity of the type of Eq.~\eqref{eq:34}, hence showing the same scaling as the voter model.

\section{Family of Voter Models \label{sec:4}}
\subsection*{The voter model}

The voter model is a paradigmatic binary-state stochastic model, with applications to physical, biological, chemical and social complex systems \cite{clsu73,holi75,cafolo09,karebe10}. It considers an ensemble of $N$ equivalent elements, also called agents, endowed with two possible states, namely, $+1$ or $-1$, frequently called the opinion. We define $n\in[0,N]$ as the total number of agents in state $+1$, so the associated magnetization $x=2n/N-1\in[-1,1]$ is a relevant quantity to study the global time evolution of the system. The extreme values $x = \pm 1$ describe consensus states, in which all agents agree, while $x=0$ corresponds to an equal coexistence of opinions. The standard voter model (VM) features a stochastic evolution for $x$ based on an imitation process, that is, the state of an agent changes by adopting the opinion of a neighbor in a lattice or a network of interactions. At the mean-field level, to be considered here, all agents are neighbors of each others (fully connected network). Consensus states are pure absorbing states of the dynamics: once the system reaches them, it cannot leave. A natural question concerns the probability density, or its moments, for the time the system needs to reach a consensus state for the first time given an initial condition. This and related questions have been partially addressed recently \cite{soanre08,vare04,magire10,lare07,rogo13,sabapa12,bacapa11,molicapa13,capa12,camupa09,iwma14,ma14} and will be reconsidered next.

In the framework of the Monte Carlo simulations, the mean field voter model can be defined as follows. At each update event, two randomly chosen agents interact and modify their opinions according to the model, i.e. one of them blindly copies the state of the other. The repetition of $N$ of such interactions computes as $1$ MCS. We keep repeating this dynamics until a steady state is reached. Consensus is the final fate of the model as far as the system size $N$ remains finite. Moreover, if $N$ is finite but large enough, the probability distribution for the global magnetization obeys the Fokker-Planck equation \eqref{eq:14} with $A=0$ and $B(x)=(1-x^2)/(2N)$ \cite{soanre08}. Our analysis of the family of voter models is restricted to finite values of $N$.

We illustrate in Fig.~\ref{fig:3} the same time distributions already addressed in the previous models. We plot again the analytical solution for the voter model [Eqs.~\eqref{eq:21} and \eqref{eq:35}] and the results coming from the simulations. Additionally, in each corresponding case we plot also the results of the WKB calculations \eqref{eq:38}--\eqref{eq:37}, which capture the expected power-law behavior. It is worth mentioning the discrepancy between the theoretical solution and the simulation results in the third panel of Fig.~\ref{fig:3}, as the very same phenomenon will also appear when analyzing other models that experience the effects of absorbing states. The small-time difference comes from the fact that we use a finite number of addends in the final solution and eventually disappears when the sum is computed with infinite terms. However, the mismatch in the tail has a different origin: the theoretical solution is a continuous approximation of the discrete process, the voter model, which is the one we actually simulate. When an absorbing state is approached, the transition rates between states tend to 0, but they do it in a different manner depending on whether a continuous or a discrete case is considered. These rates become equal when the distance between adjacent states become 0; otherwise the discrete rates are smaller and the dynamics is trapped. The consequence of this, as can be observed in Fig.~\ref{fig:3}III, is that simulations take longer to decay than the analytical, continuous solution. We will give quantitative arguments at a further point to better understand the discrepancy.

\begin{figure}
 \centering
 \includegraphics[width=0.8\linewidth]{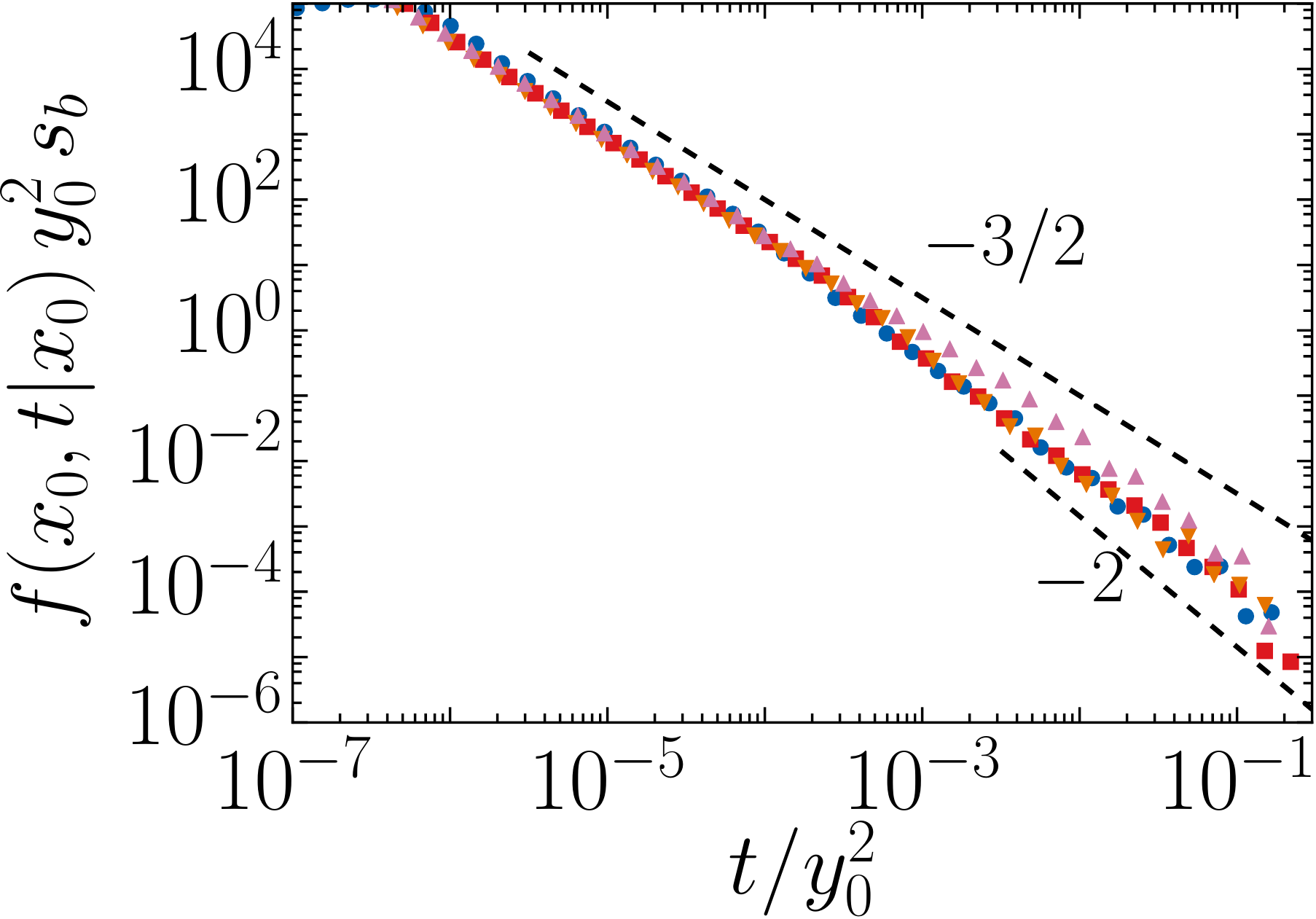}
 \caption{Return distributions for the voter model with a system size $N=10^{4}$. Symbols come from Monte Carlo simulations, varying the initial condition: $x_0 = 0.998 $ (blue circle), $0.995$ (red square), $0.99$ (orange down triangle), and $0.9$ (pink up triangle) for $N=10^{4}$. Dashed lines correspond to the power laws of exponent $ -2 $ and $ -3/2 $, to guide the eye. For the sake of clearness, the distributions have been multiplied by different factors $s_b$ and the exact and approximate theoretical lines have been dropped.}
 \label{fig:4}
\end{figure}

As discussed at the end of the preceding section, we can have a crossover effect between the power-law of exponent $ -2 $ (due to the absorbing state) and the exponent $ -3/2 $ (free exploration of the interval). This occurs when there is no dominant effect of one of the above elements over the other. To illustrate this effect, we show in Fig. \ref{fig:4} the FR distribution for the voter model, varying the initial condition $ x_0 $, i.e., modifying the influence of the singularity on the stochastic evolution of the variable $ x $. We see that if the starting point is close to the boundary, the exponent $ -2 $ dominates in the distribution; however if we place $ x_0 $ away from the singularity, the exponent $ -3/2 $ appears for small times. The further the starting point is from the singularity, the longer the $ -3/2 $ decay dominates.

\subsection*{Noisy voter models}

All scenarios studied up until now correspond to situations where the singularities of the $ \Delta(x) $ function are accessible to the system, i.e., they are within the valid range of $ x $. However, when noise is introduced in the voter model (understood as spontaneous opinion flips), $ \Delta(x) $ is still singular but the singularities fall outside the interval $ [-1,1] $. Put otherwise, the system could display absorbing states, but they are practically inaccessible. In this case the consensus states still exist, but they are no longer absorbing: due to noise, the system can leave them. We now study the effects of this type of noise on the FP and FR distributions.

We focus on a family of models that, besides the copying mechanism based on interactions among nodes, include too a mechanism of opinion change intrinsic to the agents, the noise. In particular, two ways of implementing such a mechanism will be considered: the voter model with global noise (VMGN) and the noisy voter model or Kirman model (KM) \cite{ki93,grma95,khsato18,catosa16,pecasato18,diegsa15}. On the one hand, in the VMGN we have two kind of events: at each update, with probability $q$ a node changes state to the state of another node chosen at random, just the standard VM; with probability $1-q$, the total magnetization of the system decreases or increases, with the same probability, an amount $2/N$  (except for the extreme values $x=\pm 1$), that is, the update is driven by pure noise at a global level. The latter is just like a random walk event in the magnetization space. For example, let us consider a scenario in which we have $ 90 $ nodes in states $ +1 $, out of $ 100 $. A noisy update of the VMGN will lead the system to a number of $ 89 $ or $ 91 $ nodes in state $+1$ with equal probability: the noise is magnetization independent. On the other hand, in the Kirman model with probability $q$ we perform a standard voter model update, whereas with probability $1-q$ a random node is selected and changes its state. The crucial difference in this model, compared to the voter model with global noise, is that the noise is magnetization dependent, that is, if we have a majority of nodes $+1$, a transition $+1 \rightarrow -1$ is more likely to be observed. In the same scenario as before, in the noisy update of the Kirman model the transition to $ 89 $ nodes in state $+1$ will occur with probability $ 0.9 $, while the other transition will occur with probability $0.1$.

When the number of agents $N$ and the typical timescale evolution of the system are large enough, the variable $x$ can be regarded as continuous and its probability density $p(x,t)$ satisfies a Fokker-Planck equation. More precisely, $p(x,t)$ satisfies Eq.~\eqref{eq:14} with the drift and diffusion terms given by
\begin{eqnarray}
 \label{eq:41}
 && A(x)=-q\frac{k}{2}x, \\
 \label{eq:42}
 && B(x)=\frac{1}{N}\left[q+\frac{1-q}{2}(1-x^2)\right],
\end{eqnarray}
where $k=0$ holds for the VMGN and $k=1$ for the KM. This two-parameter model also includes all the other models studied so far with the proper choice of the parameters $q$ and $k$ (see Table~\ref{tab:1}). Note that in the case of the voter model with global noise, upon varying $q$ from $0$ to $1$ we interpolate between the voter model and the random walk. This is no longer true if $ k \neq 0 $ because a drift term appears, so the random walk cannot be recovered.

\begin{table}[]
 \includegraphics[width=\linewidth]{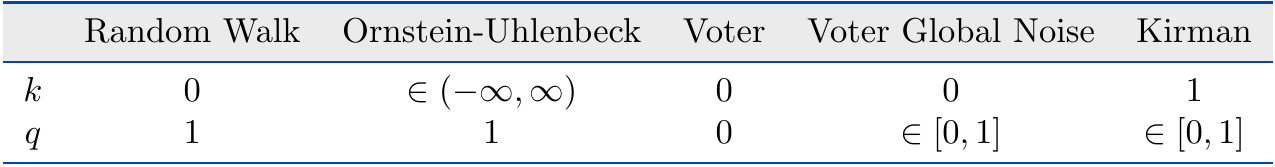}
 \caption{Reduction of the general model ( Eqs. \eqref{eq:41} and \eqref{eq:42}) to specific models for different values of the parameters $q$ and $k$.}
 \label{tab:1}
\end{table}

\subsection*{Analytical results}

As discussed in Sec. \ref{sec:2}, in order to obtain the FP distributions we first have to solve the corresponding eigenvalue problem, that is, Eq.~\eqref{eq:17} with the drift and diffusion coefficients given by \eqref{eq:41} and \eqref{eq:42}. The eigenvalue equation to be solved is
\begin{equation}
 \label{eq:43}
 \begin{split}
 & \frac{1}{2N}\frac{d^2}{dx^2}\left\{\left[q+\frac{1-q}{2}(1-x^2)\right]X(x)\right\}\\
 & \qquad +q\frac{k}{2}\frac{d}{dx}\left[xX(x)\right]+\lambda X(x)=0.
 \end{split}
\end{equation}
If we seek a solution of the form $X(x)=(1-z^2)^{\varepsilon}Z(z)$ with $z=\sqrt{\frac{1-q}{1+q}}x$ and $\varepsilon$ an exponent to be determined, Eq.~\eqref{eq:43} can be transformed into an associated Legendre equation
\begin{equation}
 \label{eq:44}
 \begin{split}
 & (1-z^2)\frac{d^2}{dz^2}Z(z)-2z\frac{d}{dz}Z(z)\\ & \qquad +\left[\nu(\nu+1)-\frac{\mu^2}{1-z^2}\right]Z(z)=0,
 \end{split}
\end{equation}
with
\begin{equation}
 \label{eq:45}
 \varepsilon=-\frac{1}{2}\left(1-Nk\frac{q}{1-q}\right),
\end{equation}
and
\begin{eqnarray}
 \label{eq:46}
 && \mu=\pm 2 \varepsilon, \\
 \label{eq:47}
 && \nu=-\frac{1}{2}\left[1\pm \sqrt{(1+4\varepsilon)^2+\frac{16N\lambda}{1-q}}\right].
\end{eqnarray}
A general solution to Eq.~\eqref{eq:43} can be constructed by means of the Ferrers \cite{ollobocl10} function of the first kind $P_\nu^{\mu}(x)$ of order $\mu$ and degree $\nu$. Selecting $\mu>0$, we have
\begin{eqnarray}
 \label{eq:48}
 \begin{split}
 X_n(x)=&\left(1+\frac{1-q}{1+q}x^2\right)^{\varepsilon}\left[A_n P_\nu^{-\mu}\left(\sqrt{\frac{1-q}{1+q}}x\right)\right.
 \\ & \qquad \left. +B_n P_\nu^{-\mu}\left(-\sqrt{\frac{1-q}{1+q}}x\right)\right],
 \end{split}
\end{eqnarray}
provided $\nu+\mu\ne -1,-2,\dots$ and $\mu-\nu\ne 0,-1,-2,\dots$; otherwise the two terms of the sum are linearly dependent. Since $P_\nu^\mu (x) = P_{-(\nu+1)}^\mu(x)$, the proposed solution accounts for the two values of $\nu$ allowed by Eq.~\eqref{eq:47}. The constants $A_n$ and $B_n$, and the eigenvalues $\lambda_n$ have to be determined by imposing the boundary conditions and the normalization of $X_n$. The boundary conditions are
\begin{eqnarray}
 \label{eq:49}
 && X_n(x_f)=0, \\
 \label{eq:50}
 && J[x_*|X_n]=0, 
\end{eqnarray}
with $x_*=-1$ if $x_0<x_f$ or $x_*=1$ if $x_0>x_f$. We could also use an absorbing boundary condition at $x_*$ [$X_n(x_*)=0$] but we find more natural the reflecting ones in the context of the voterlike models.

\subsection*{Effects of global noise and absorbing states}

We can increase or decrease the effect of the absorbing state by modifying the amount of noise in the system. In general, since the VMGN has no drift, the diffusion term of Eq.~\eqref{eq:42} gives the absorbing states, located at $x=\pm\sqrt{\frac{1+q}{1-q}}$, which are at the border of the available values of the magnetization for a noise parameter $q=0$ (VM), and fall outside the interval $(-1,1)$ for $q>0$, moving to infinity for $q=1$ (RW). We expect that in the regime $ q \gtrsim 0 $, the system experiences the effects of the absorbing states, although they are inaccessible. In the following, we study the influence of noise on the FP and FR distributions.

\begin{figure*}
 \centering
 \includegraphics[width=0.24\linewidth]{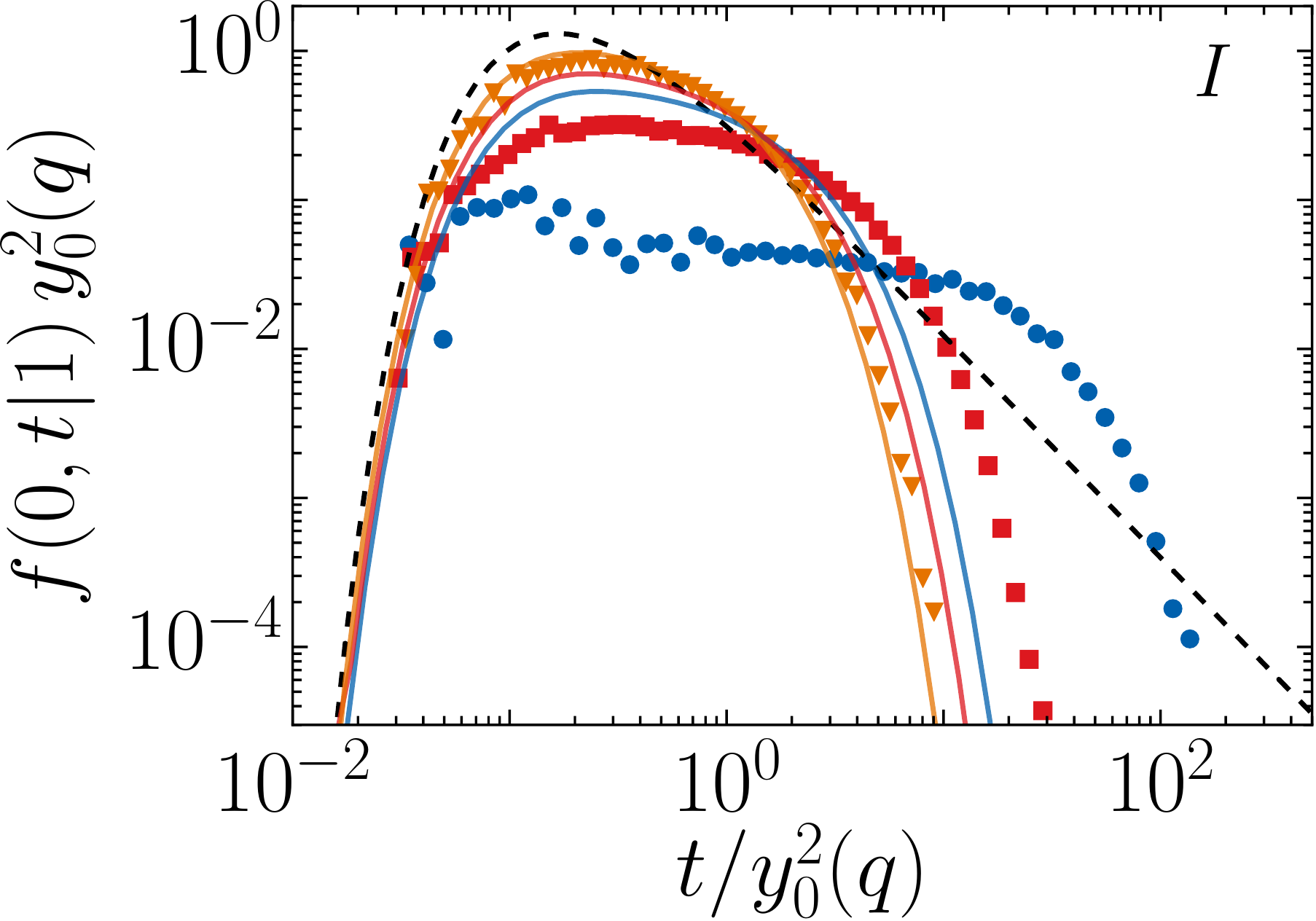}
 \includegraphics[width=0.24\linewidth]{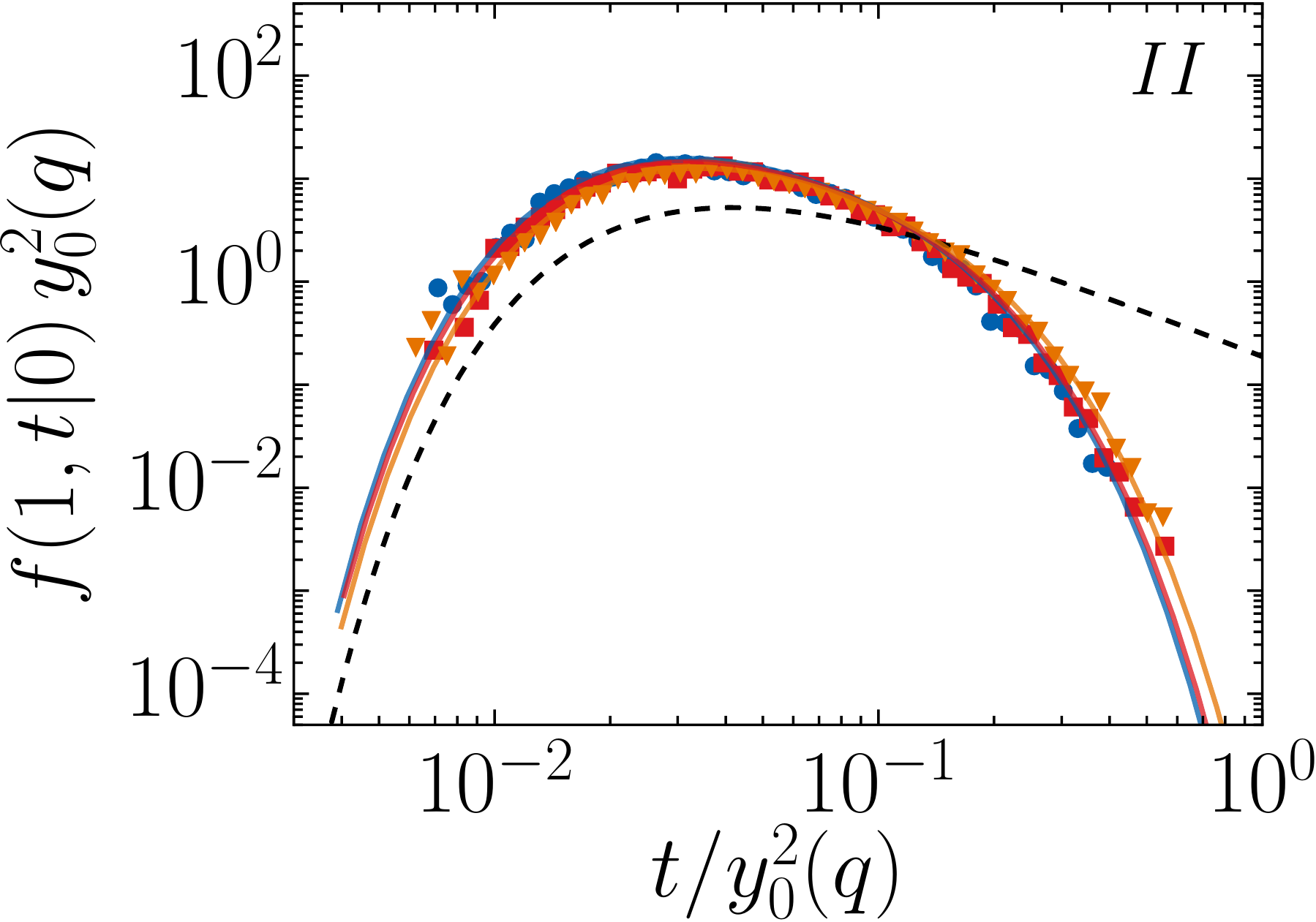}
 \includegraphics[width=0.24\linewidth]{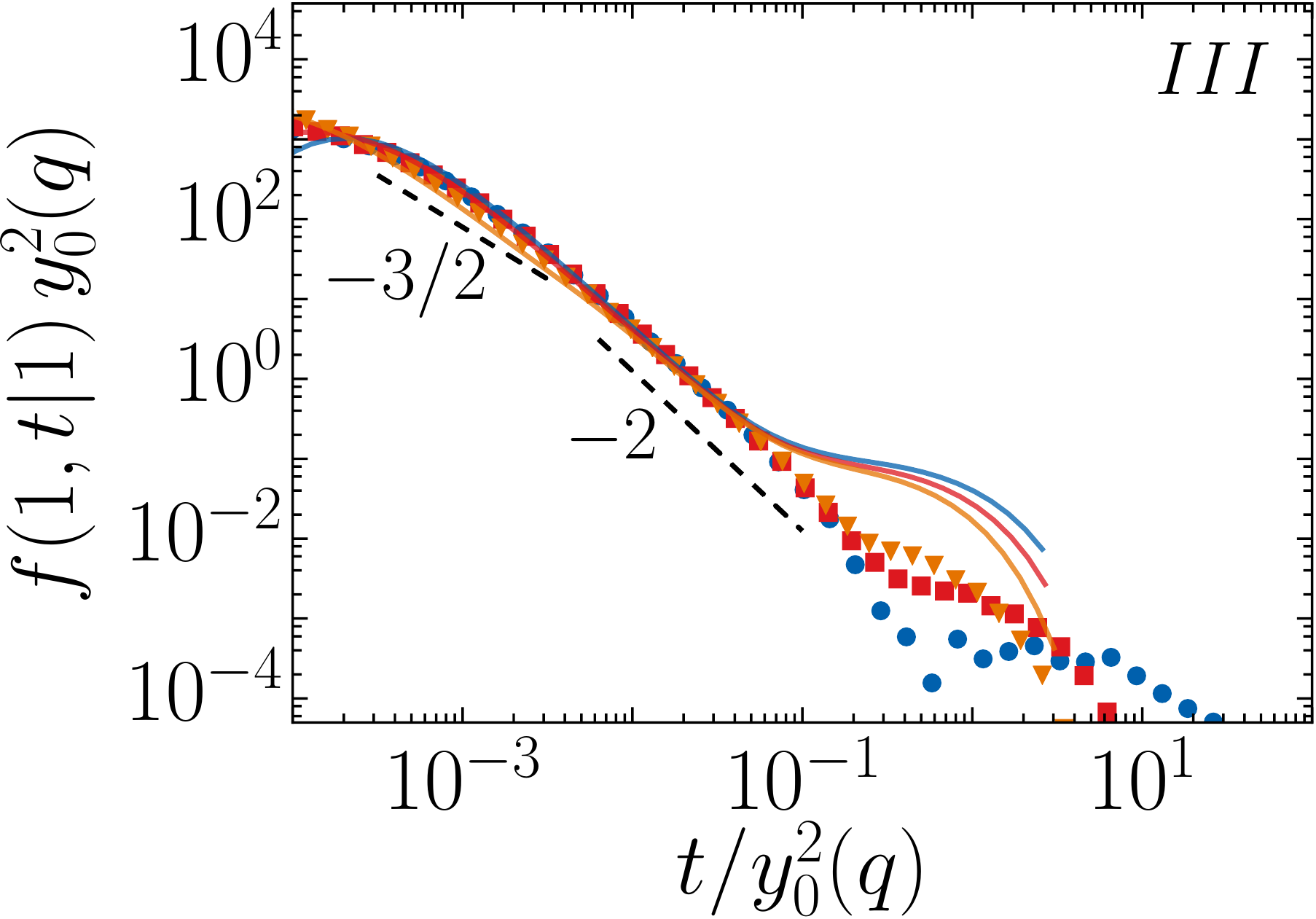}
 \includegraphics[width=0.24\linewidth]{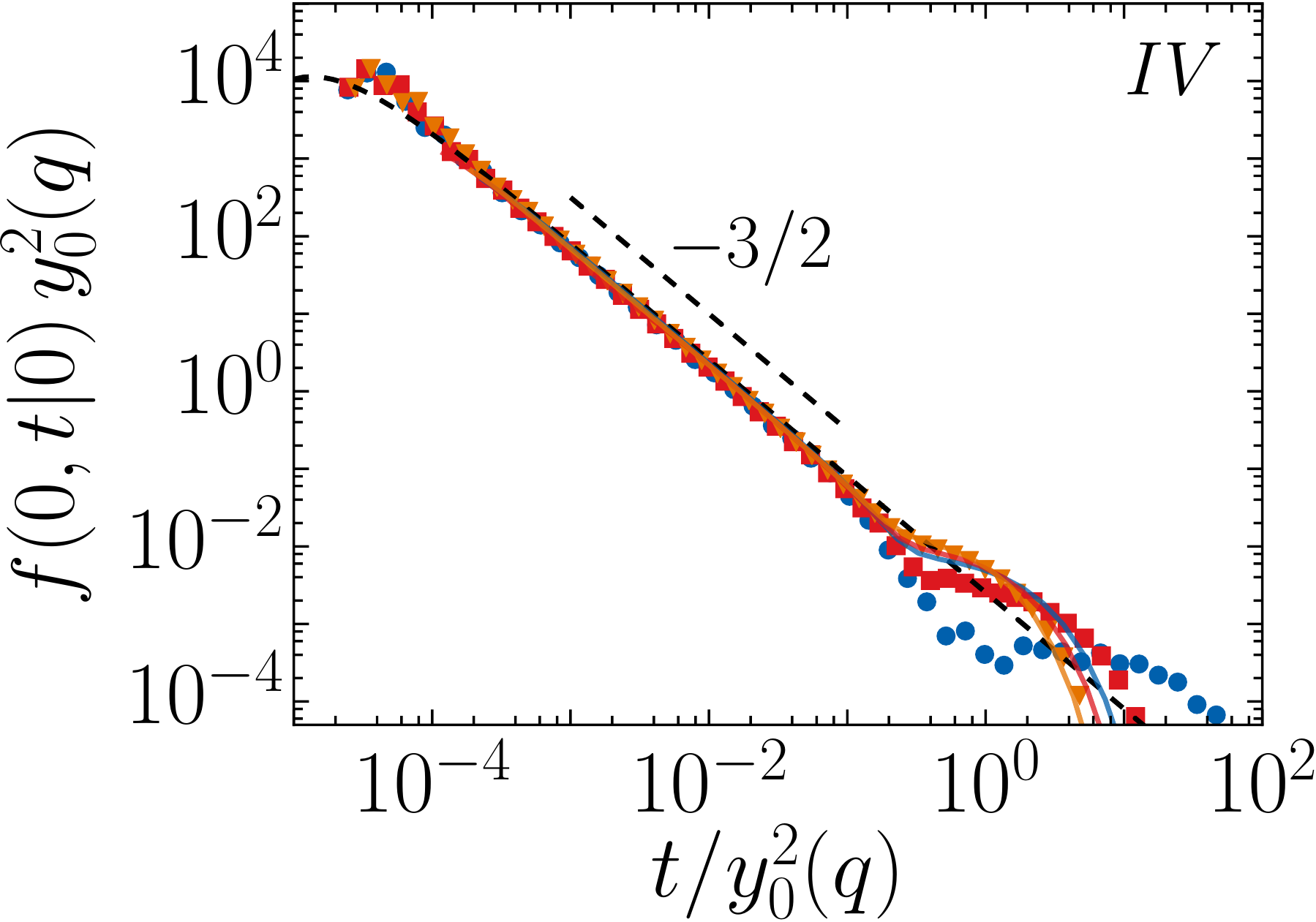}
 \caption{First-passage and first-return distributions for the voter model with global noise ($k=0$) with $q=10^{-4}$ (blue circle), $10^{-3}$ (red square), and $10^{-2}$ (orange triangle). Symbols are from Monte Carlo simulations for $N=200$, solid lines are from the exact theory of Eq.~\eqref{eq:21} (with 10 to 50 addends), and dashed lines from Eq.~\eqref{eq:27}.}
 \label{fig:5}
\end{figure*}

We analyze in Fig. \ref{fig:5} the dependence on the noise parameter $ q $ of the distributions of the four cases sketched in Fig. \ref{fig:2a} for the voter model with global noise. Figures~\ref{fig:5}I and \ref{fig:5}II show that the FP distributions from the origin (coexistence) to the borders (consensus) and vice versa depend very weakly on the values of $q$, their shape being very similar to the ones for the RW and VM. This effect is surprising, especially for the case of the FP distribution from consensus to coexistence [Fig. \ref{fig:5}I], when $q$ is very small, since some effect from the absorbing state should be expected. In fact, this is the case for the numerical simulations, where a plateau appears. This plateau disappears if the number of nodes increases.

The return distributions to consensus (case III) are shown in Fig. \ref{fig:5}III. There are three main parts: a short-time part, a second part that exhibits two power-law decays, the first one with exponent $-3/2$ and a second one with exponent $-2$, and a last part with a plateau followed by an exponential decay. As already analyzed for the case of the VM (Fig. \ref{fig:4}), the power law of exponent $-3/2$ is a consequence of the first exploration of regions far from the absorbing state, while the one with exponent $-2$ appears as a result of the influence of the absorbing point near $x=\pm 1$. This scenario is the same for almost all values of $q$, although we observe that the width of the region with exponent $-2$ narrows when increasing $q$. The plateau for long times also appears due to the presence of the absorbing states near the boundary of the interval.

Finally, Fig. \ref{fig:5}IV shows the return distribution to coexistence of opinions or, equivalently, the FP from $x_0=0$ to $x_f=0$ (case IV). For a wide time window the distribution displays a power law of exponent $-3/2$, as for the RW. The situation resembles the one for the VM, for similar initial and final points, something expected since there is no important contribution to this quantity from the absorbing state for intermediate times. The role of the absorbing state is markedly present in the tails of the distributions, where a plateau appears as a consequence of the trapping of the system at the boundaries, which lasts longer as $q$ approaches $0$. When there is no noise, $q=0$, one recovers the voter model and, accordingly, the plateaus disappear since there is no way of leaving the consensus state.

The difference between the Monte Carlo simulations (symbols) and the theoretical predictions (lines), in both Figs.~\ref{fig:5} and \ref{fig:6}, lies in the failure of the continuous limit ($N$ is not big enough), that is, in the difference between the simulations and the Fokker-Planck equation mainly near the absorbing points. This is easy to understand if we estimate, as an example, the typical time the system needs to go from one point to an adjacent one $x\to x\pm 2/N$. In the discrete case, it is $\tau_d\sim [q+(1-q)(1-x^2)/2]^{-1}$, which is of the same order as for the continuous case $\tau_c\sim |N\int_{x}^{x\pm 2/N} dx[q+(1-q)(1-x^2)/2]|^{-1}$, except for $x=\pm 1$, where we have $\tau_d\sim q^{-1}$ and $\tau_c\sim [q+1/(2N)]^{-1}$. Hence, when the extreme points are being explored by the system and the level of noise is $q\lesssim 1/(2N)$, the numerical (discrete) distribution is a factor of time of the order of $q^{-1}-2N$ slower than the analytic one. For $N\to \infty$, it is $\tau_d\sim \tau_c$ and the discrepancies disappear.

\subsection*{Effect of different phases}

An important conclusion from the analysis of the preceding subsection regarding the voter model with global noise can be inferred. Namely, the FP and FR distributions depend weakly on the parameter $q$, provided the time is measured in units of $y_0^2(q)$. This is related to the fact that the system is always in a bimodal phase, that is, the steady-state solution of the Fokker-Planck equation for the VMGN is always a convex function with two maxima at $x=\pm 1$, regardless of the values of $q\in (0,1)$. The Kirman model, by introducing a drift term proportional to $x$ that pushes the system toward the origin, allows the opportunity to go beyond the VMGN by allowing the system to be also in a unimodal phase with $p_{st}(x)$ peaked around $x=0$ and where $x=\pm 1$ are global minima, i.e., the least probable values of the steady state magnetization. The transition from the unimodal to the bimodal phase occurs at $q_c=1/(N+1)$ \cite{ki93,catosa15}.

\begin{figure*}
 \centering
 \includegraphics[width=0.24\linewidth]{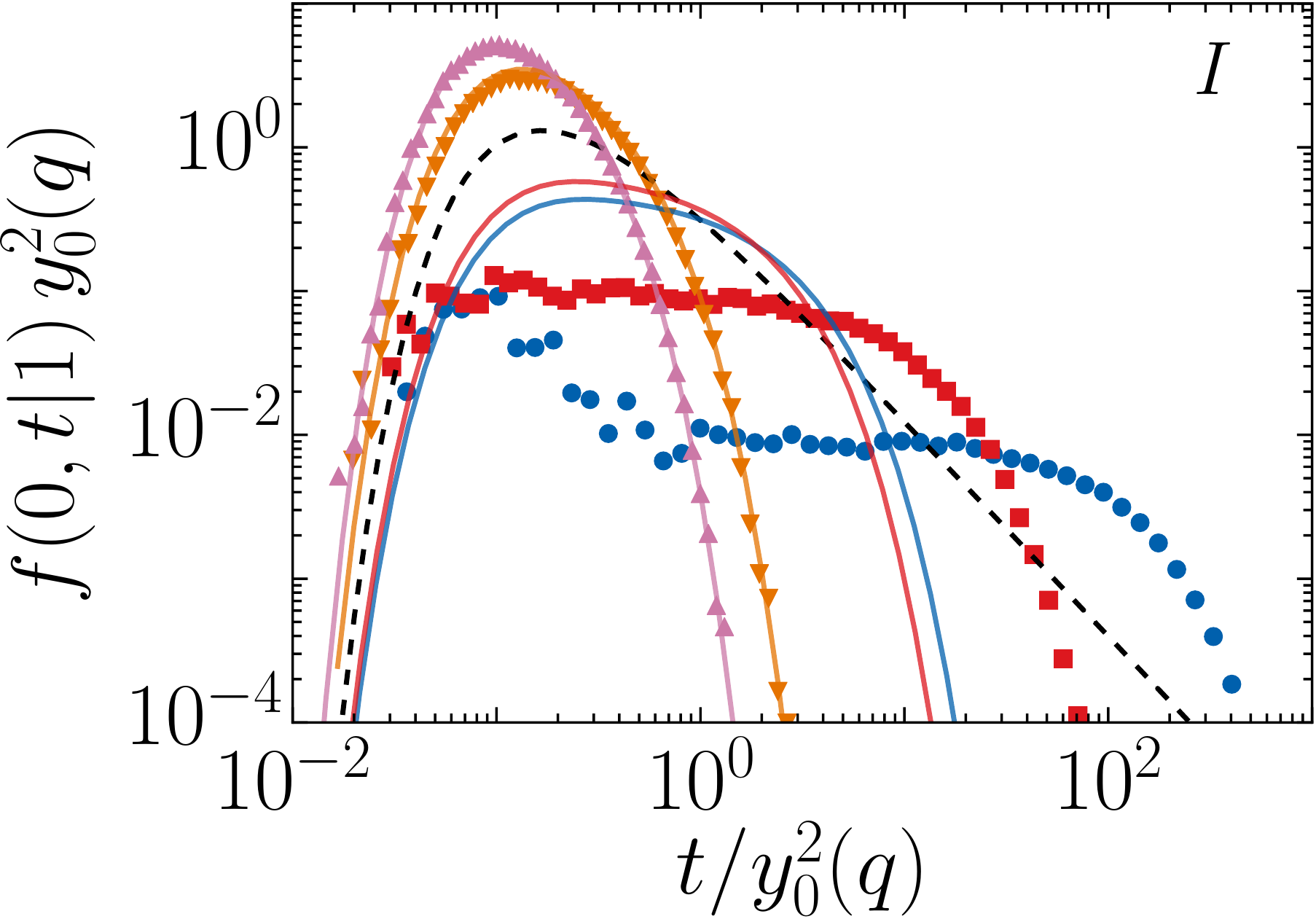}
 \includegraphics[width=0.24\linewidth]{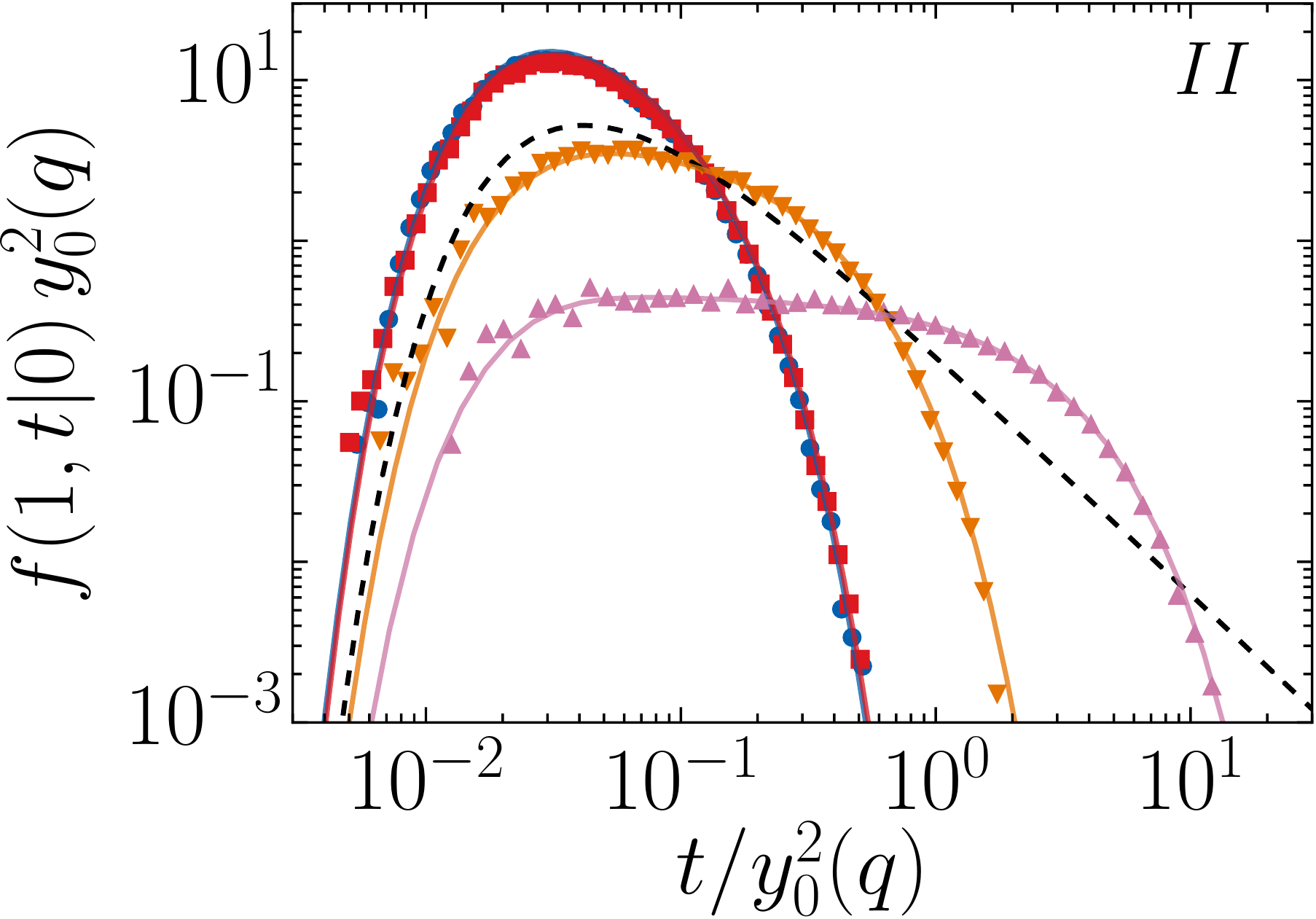}
 \includegraphics[width=0.24\linewidth]{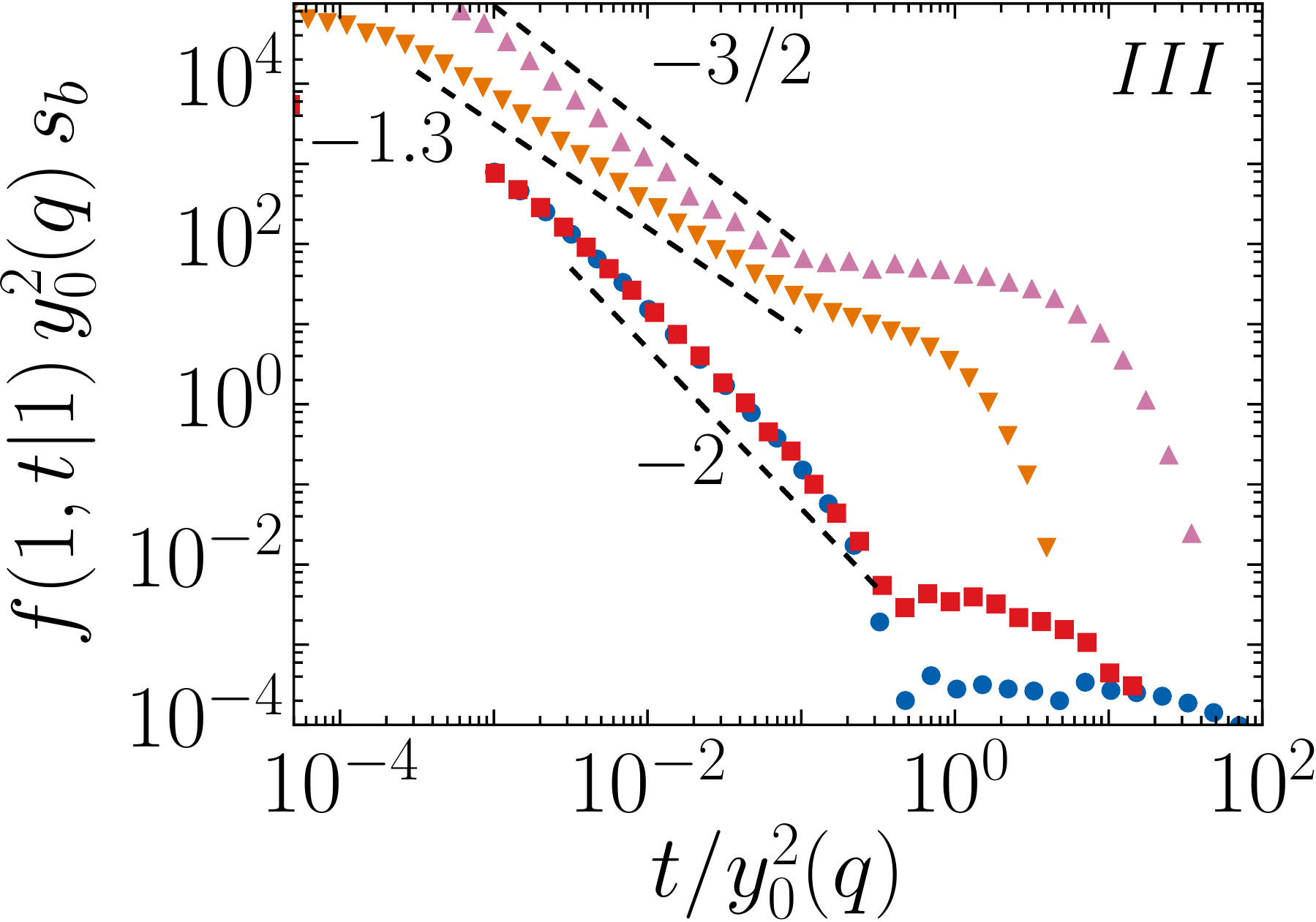}
 \includegraphics[width=0.24\linewidth]{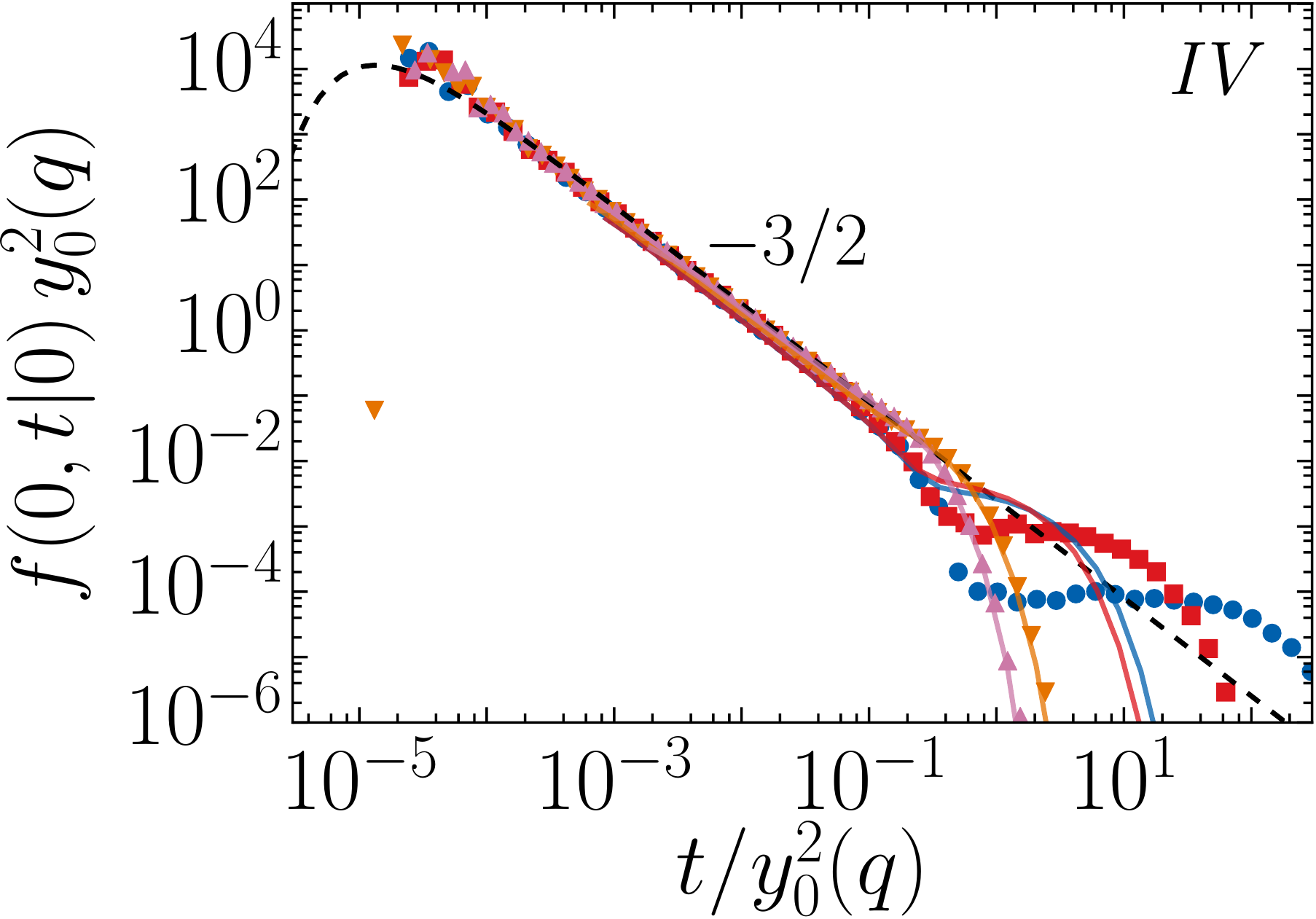}
 \caption{First-passage and first-return distributions for the Kirman model ($k=1$) with $q=10^{-5}$ (blue circle), $10^{-4}$ (red square), $q_c=5\times 10^{-3}$ (orange down triangle), and $10^{-2}$ (pink up triangle). Symbols are from Monte Carlo simulations for $N=200$, solid lines are from the exact theory (with 10 to 50 addends), and dashed lines from Eq.~\eqref{eq:27}. Data of the third plot have been multiplied by different factors ($s_b$) in order to better appreciate the different power laws.}
 \label{fig:6}
\end{figure*}

In Fig. \ref{fig:6} we have plotted for the Kirman model the same magnitudes as in Fig. \ref{fig:5}. We observe that for $q<q_c$, i.e. in the bimodal phase, the FP and FR distributions of the Kirman model behave as that of the voter model with global noise. Important differences appear when $q>q_c$:
\begin{itemize}
\item The FP distribution from coexistence to consensus develops a plateau at intermediate times [case II, Fig. \ref{fig:6}II].
\item The FR distribution to a consensus state loses its power law with exponent $-2$ [case III, Fig. \ref{fig:6}III].
\item The FR distributions to the origin lose their long-time plateau [case IV, Fig. \ref{fig:6}IV].
\end{itemize}
More interestingly, at $q=q_c$ the return distribution to a consensus state develops a power-law decay of exponent $\sim -1.3$, which turns out to be independent of the number of agents $N$.

\subsection*{Inferring the phase from the first-passage distributions}

The value of the noise parameter $ q $ determines the phase of the system, i.e., the unimodal or bimodal structure of the probability density function of the magnetization in the steady state. However, we now show that the phase can be inferred from properties of the FR distribution too.
In fact, there are two features of this distribution that provide information about the phase of the system. One option is to look at the FR to any of the consensus states [case III, Fig. \ref{fig:6}III], the exponent of the intermediate power laws is different for the two phases. The exponent in the bimodal phase is $ -2 $, whereas an abrupt change occurs around the critical point, being the exponent $ \sim -1.3 $. Once we enter the unimodal phase, we recover the exponent $ -3/2 $. The other option is to investigate the FR to coexistence [case IV, Fig. \ref{fig:6}IV], where the presence or not of a final plateau indicates the bimodal or unimodal character of the phase, respectively. Let us analyze this second option, the FR distribution to the center of the interval. The absorbing boundary condition enforces the eigenfunctions \eqref{eq:48} to satisfy $X_n(0)=0$, which implies $A_n+B_n=0$. In addition, owing to the reflecting nature of the boundary, we have $J[1|X_n]=0$. The latter involves $X_n(1)$ and its derivative at $x=1$, namely, expressions with $P_\nu^{-\mu}\left(\pm\sqrt{\frac{1-q}{1+q}}\right)$ and $P_{\nu+1}^{-\mu}\left(\pm\sqrt{\frac{1-q}{1+q}}\right)$. Setting the value of the noise to $q\sim 1/N$, i.e. close the critical point, the arguments of the Ferrers functions are close to $\pm 1$. In this limit, we have 
\begin{equation} P_\nu^{-\mu}\left(\sqrt{\frac{1-q}{1+q}}\right)\sim \frac{1}{\Gamma\left(1+\mu\right)}\left(\frac{q}{2}\right)^{\frac{\mu}{2}}
\end{equation}
and 
\begin{equation}
P_\nu^{-\mu}\left(-\sqrt{\frac{1-q}{1+q}}\right)\sim -\frac{\Gamma(\nu-\mu+1) \sin[(\kappa-\mu)\pi]}{\Gamma(\nu+\mu+1)\sin(\mu\pi)}\frac{1}{\Gamma(1+\mu)}\left(\frac{q}{2}\right)^{\frac{-\mu}{2}},
\end{equation}
which implies for, $\mu>0$, $P_\nu^{-\mu}\left(\sqrt{\frac{1-q}{1+q}}\right)\ll P_\nu^{-\mu}\left(-\sqrt{\frac{1-q}{1+q}}\right)$. The boundary conditions impose the equation
\begin{equation}
 \label{eq:51}
 \left[\nu-\frac{2(1+\varepsilon)-1+(\mu-1)\sqrt{\frac{1+q}{1-q}}}{\sqrt{\frac{1+q}{1-q}}-1}\right] \sin[(\nu-\mu)\pi] \simeq 0,
\end{equation}
from which we obtain an approximate expression for the eigenvalues
\begin{equation}
 \label{eq:52}
 \lambda_0\simeq \left\{
 \begin{split}
 & {\scriptstyle o(q) \ \text{ if } \ Nk\frac{q}{1-q}\le 1}, \\
 & {\scriptstyle \frac{\sqrt{1-q^2}(1-q+\sqrt{1-q^2})\varepsilon}{2Nq}\left[\frac{\sqrt{1-q}+\sqrt{1+q}}{2}\varepsilon-1\right] \ \text{ if }\ Nk\frac{q}{1-q}>1,}
 \end{split}
 \right.
\end{equation}
and
\begin{equation}
 \label{eq:53}
 \lambda_n\simeq \frac{1-q}{4N}\left\lbrace n(n+1)+2\left[|\varepsilon|(2n+1)-\varepsilon\right]\right\rbrace,
\end{equation}
for $n=1,2,\dots$ We recover the result of the voter model [$q=0$, see Eq.~\eqref{eq:35}], after changing $n$ to $n-1$ (since in the VM the mode associated with $\lambda=0$ was disregarded). Interestingly, the smallest eigenvalue is approximately zero (much less than $q$) in the bimodal phase and different from zero in the unimodal phase. That means that for $N\gg 1$ the system gets trapped close to the borders only in the bimodal phase. For finite $N$, though, the FP distribution develops a plateau in the bimodal phase.

\section{Discussion, conclusions, and future perspectives \label{sec:5}}

The problem of characterizing the FP (first-passage) and the FR (first-return distributions) for the 1D Fokker-Planck equation in bounded domains, with generic position-dependent diffusion and drift terms, has been tackled. By means of the WKB approximation, we demonstrate that both functions may exhibit general features whose properties depend essentially on the eventual presence of absorbing states (where the diffusion coefficient vanishes or the drift term diverges towards this state) in which the dynamics is trapped. Among all possible cases, we have focused on two general situations: One is the class of models with the same general characteristics as the random walker (RW), and the other the class of models with the general characteristics of the voter model (VM). When there is no absorbing state affecting the system dynamics (the RW case), the first-passage and the first-return distributions for small and intermediate times are given by Eq.~\eqref{eq:27}, which can lead to a power-law decay with exponent $-3/2$. In the long-time limit, the decay of the FP and FR distributions is always exponential with a timescale strongly dependent on the diffusion and drift terms, compromising for finite domains the possible appearance of the power law at intermediate times. If an absorbing state exists, the RW behaviour may break down with the eventual appearance of new exponents. If the diffusion coefficient vanishes linearly at one point accessible to the system, or more generally if the quantity at Eq.~\eqref{eq:28b} diverges as in Eq.~\eqref{eq:34}, the behavior of the FP and FR distributions is that of the class of model of the VM. For these models, if the initial and final states for computing the FP and FR distributions are far enough from the absorbing sates, the behaviour of the random walk still prevails. In the other extreme case where the initial and final states are close to the absorbing state, a new power law at intermediate times with exponent $-2$ is found.


To check our theoretical predictions, we have discussed five models of increasing complexity. As examples of systems without absorbing states we have explored \textit{(i)} the well-known random (Brownian) walker, characterized by constant diffusion and no drift, and \textit{(ii)} the classical Ornstein-Uhlenbeck process, with a linear drift and constant diffusion. With these two models we aimed at evaluating the influence of a drift on the FP and FR distributions. The RW exponent prevails for almost all drifts considered defining a class of models with power-law decay with exponent $-3/2$. However, these results are different from those obtained in semi infinite domains where a nonalgebraic decay is observed \cite{mascwe12}. The three remaining models have in common the existence of an absorbing state in the dynamics, which eventually can be outside the considered domain. They are \textit{(iii)} the voter model (VM), that has a space-dependent diffusion but lacks drift and where the two boundaries are natural absorbing states of the dynamics; \textit{(iv)} the voter model with global noise, which has no drift, but has no absorbing states because of the effects of noise acting at a global level (this model depends on a continuous parameter that interpolates between the RW and VM); and \textit{(v)} the Kirman model, which displays drift and diffusion owing to a magnetization-dependent noise and has no absorbing states. The interest now has been the evaluation of the effect of the absorbing state on the FP and FR distributions. We demonstrate, both analytically and numerically, that the two general behaviours of the RW and VM classes of models can appear separately or even together. Beyond an intermediate decay, the system may exhibit plateaus in the FP and FR times due to the trapping nature of the absorbing states, a feature that has been used to infer the phase, either unimodal or bimodal, of the system dynamics.

The generality of the 1D Fokker-Planck equation together with the weak conditions required on the drift and diffusion coefficients give a broad applicability to our results. This generality relies on the possibility of expressing the FP and FR distributions as a superposition of modes, on the contribution of many of these modes, and also on the possibility of approximating this superposition  to that of the RW if there is no absorbing state, or that of the VM (with the appropriate absorbing state).


There are some open questions that fall beyond the scope of this work. A main point concerns the nature of the absorbing state. In the present paper we have focused on those absorbing states characterized by a linearly vanishing diffusion coefficient. Other physical systems might have absorbing states with another functional dependence on the state variable, but still our theory can be applied to them. It would certainly be interesting to study the same distributions in higher dimensions of the Fokker-Planck equation, particularly $ d = 2 $, which is critical for the random walk. Finally, real physical systems are highly correlated, display memory, and their statistics, for example, in the jump step-size or in the waiting times, are non-Poissonian \cite{bolatose14}. It is an interesting task for future research to study the impact of these elements on our theoretical predictions.

\section*{Acknowledgements}

Partial financial support has been received from the Agencia Estatal de Investigaci\'on (AEI, Spain) and Fondo Europeo de Desarrollo Regional under project ESOTECOS FIS2015-63628-C2-2-R (MINECO/AEI/FEDER,UE).

\appendix

\section{First-passage distributions for generic dynamics and dimension\label{app:1}}

We devote this appendix to formally define time quantities such as first-passage and first-return distributions, the relations among them, and the way they can be obtained from the probability density of the system. We consider the general situation of a $d$-dimensional real and continuous stochastic variable $\mathbf X(t)\in\mathbb R^d$ whose probability density $p(\mathbf x,t)$ is taken as known. Note that one realization of $\mathbf X(t)$ can be viewed as a trajectory, understood as a (measurable) subset of $\mathbb R^d$. Thereby, the probability density $p(\mathbf x,t)$ is a measure of the proportion of associated trajectories for which the stochastic variable takes the value $\mathbf x$ at time $t$. The intrinsic dynamics of $\mathbf X(t)$ will not be specified, although some restrictions will be assumed later.

\subsection*{Definitions and relations}

\begin{figure}[!h]
 \centering
 \includegraphics[width=.95\linewidth]{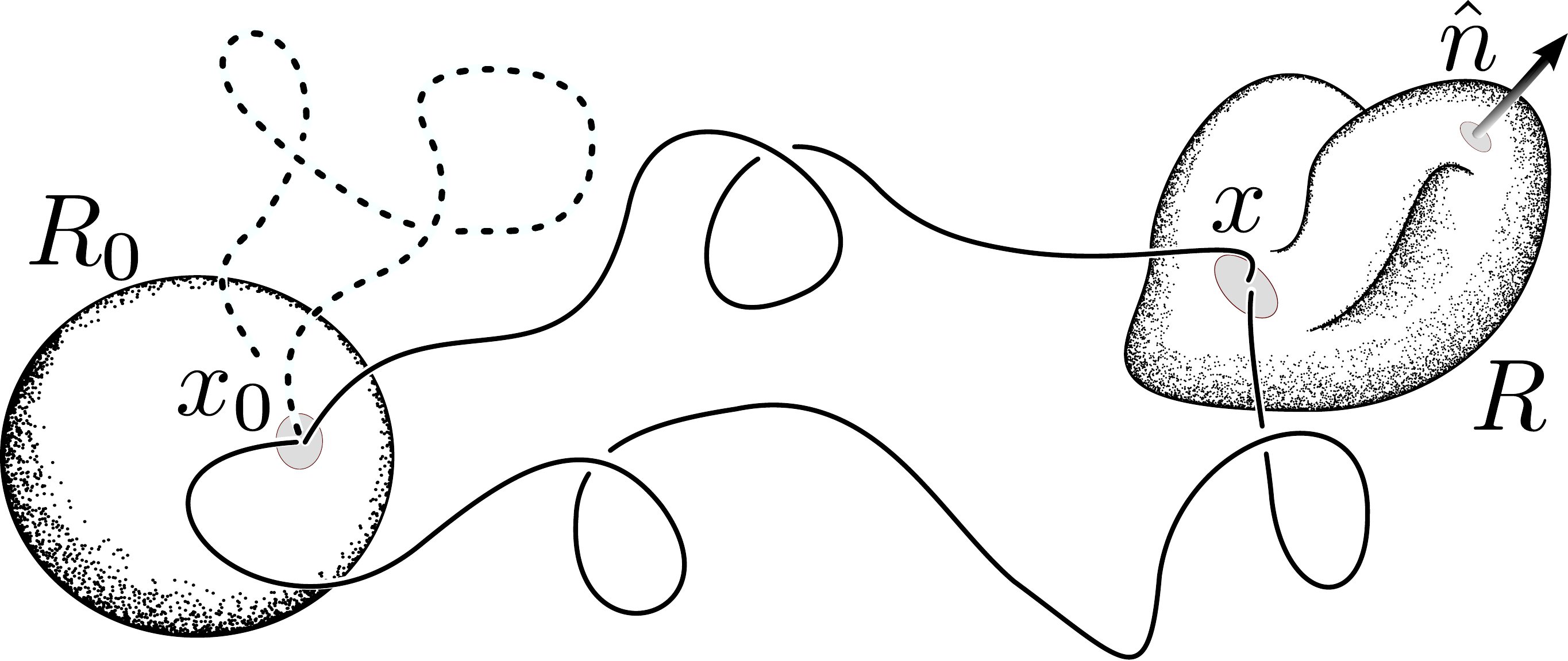}
 \caption{Schematic representation of three trajectories. Two of them (solid lines) contribute to the first-passage distribution, as it is introduced in equation \eqref{eq:1}, from point $\mathbf x_0$ to region $R$. The remaining one contributes to the first-return distribution to region $R_0$, as defined in equation \eqref{eq:7}.}
 \label{fig:1}
\end{figure}

Consider a closed region $R$ of $\mathbb R^d$ with a smooth boundary $\partial R$ and a point $\mathbf x_0\notin R$. We define the \textbf{first-passage distribution} $f(\mathbf x,R,t|\mathbf x_0,t_0)$ from point $\mathbf x_0$ to region $R$, with $\mathbf x\in\partial R$ being the first-contact point of $R$, so that
\begin{eqnarray}
 \nonumber
 && f(\mathbf x,R,t|\mathbf x_0,t_0)dS dt \equiv \quad \text{probability of $\mathbf X(t)$ to be for the} \\
 \nonumber
 && \text{\quad first time in $R$, taking a value near point $\mathbf x$ at a time} \\
 \label{eq:1}
 && \text{\quad close to $t$, provided } \mathbf X(t_0)=\mathbf x_0.
\end{eqnarray}
By ``near point $\mathbf x$'' we mean inside a ball in $\partial R$ of center $\mathbf x$ and $d-1$ area $dS$ and by ``close to time $t$'' we mean at a time between $t$ and $t+dt$. For $d=1$ there is no need to consider the ball around $x$ and we should set $dS=1$ in \eqref{eq:1}. In an intuitive way, the distribution of Eq.~\eqref{eq:1} accounts for the fraction of trajectories that start at point $\mathbf x_0$ at time $t_0$ and after a time $t-t_0$ reach for the first time region $R$ at the point $\mathbf x\in\partial R$ (see Fig. \ref{fig:1}). Hence, the stochastic process $\mathbf X(t)$ is allowed to revisit $\mathbf x_0$ after $t$.

The FP distribution of Eq.~\eqref{eq:1} is a fundamental quantity from which we can infer others, more common ones. In particular, the first-passage distribution from point $\mathbf x_0$ to the point $\mathbf x\ne \mathbf x_0$ is
\begin{equation}
 \label{eq:2}
 f(\mathbf x,t|\mathbf x_0,t_0)=\lim_{S_R\to 0} f(\mathbf x,R,t|\mathbf x_0,t_0); \qquad \mathbf x\in R,
\end{equation}
where $S_R$ is the ``area'' of $\partial R$ and the limit is assumed to exist. When the first-passage time is towards a whole region regardless of the first point reached, one can write it as
\begin{equation}
 \label{eq:3}
 f(R,t|\mathbf x_0,t_0)=\int_{\partial R} dS\ f(\mathbf x,R,t|\mathbf x_0,t_0).
\end{equation}
If, on the contrary, the trajectory departs from a region $R_0$ towards a point $\mathbf x\notin R_0$, the first-passage distribution is, in this case,
\begin{equation}
 \label{eq:4}
 f(\mathbf x,t|R_0,t_0)=\frac{\int_{R_0} d\mathbf x_0 \ f(\mathbf x,t|\mathbf x_0,t_0) p(\mathbf x_0,t_0)}{\int_{R_0}d\mathbf x_0\ p(\mathbf x_0,t_0)},
\end{equation}
where the relation \eqref{eq:2} is needed and $p(\mathbf x_0,t_0)$ is the initial probability density. Finally, when the transition is between two disjoint regions of the space, $R_0$ to $R$, the FP distribution is
 \begin{equation}
 \label{eq:5}
 f(R,t|R_0,t_0)=\frac{\int_{R_0} d\mathbf x_0 \ f(R,t|\mathbf x_0,t_0) p(\mathbf x_0,t_0)}{\int_{R_0}d\mathbf x_0\ p(\mathbf x_0,t_0)},
 \end{equation}
where now Eq.~\eqref{eq:3} is needed.

A common feature of all the definitions \eqref{eq:1}--\eqref{eq:5} is that the stochastic variable $\mathbf X(t)$ does not need not to leave $R_0$ or $\mathbf x_0$ immediately after $ t_0 $, but it can stay there for some time. The first-passage distributions presented so far do not provide any direct information about the distribution for the time spent going from one place to another, i.e., the time while the system is actually ``traveling''. In order to account for the latter case, we introduce the concept of \textbf{first-transition distributions}, denoted by $f_t$, which can be simply defined as in Eqs. \eqref{eq:1}--\eqref{eq:5} with the additional condition $\mathbf X(t_0^+)\ne \mathbf x_0$.

For the case of the departure and arrival regions being the same ones (or joint regions), we consider the \textbf{first-return distributions}. Assuming that $\mathbf X(t)$ is a Markov process and it returns to a point $\mathbf x_0$, the first-return distribution is
\begin{equation}
 \label{eq:6}
 \begin{split}
 f(\mathbf x_0,t|\mathbf x_0,t_0)=\int_{\mathbb R^d-\{\mathbf x_0\}} d\mathbf x \int d\tau \ f(\mathbf x_0,t|\mathbf x,\tau) & \\
 \times f(\mathbf x,\mathbb R^d-\{\mathbf x_0\},\tau|\mathbf x_0,t_0)&.
 \end{split}
\end{equation}
When the stochastic process goes back to a region $R_0$, the first-return distribution is then
\begin{equation}
 \label{eq:7}
 \begin{split}
 f(R_0,t|R_0,t_0)=\int_{\mathbb R^d-R_0} d\mathbf x \int d\tau \ f(R_0,t|\mathbf x,\tau)& \\
 \times f(\mathbf x,\mathbb R^d-R_0,\tau|R_0,t_0)&.
 \end{split}
\end{equation}

We can also consider, as above, the first-return distribution disregarding the time the system stays at the initial point (or region). In order to compute this distribution we only need to replace $f$ on the right-hand sides of Eqs. \eqref{eq:6} and \eqref{eq:7} by $f_t$. We identify this function as the \textbf{residence distribution} at $\mathbb R^d-R_0$, that is, the probability density for the time $t-t_0$ the system spends in region $\mathbb R^d-R_0$, with $f_t(R_0,t|R_0,t_0)$.

\subsection*{Relation between the first-passage distribution and the probability density function}

We can find at least two different ways of relating the time distribution functions that we have just defined to the probability density $p(\mathbf x,t)$ of the stochastic process $\mathbf X(t)$. The one used in \cite{re01} begins with discrete space and time, then is generalized to the continuum case, and uses the Laplace transform. However, the resulting relation turns out to be valid only for the random walk, or more generally if the trajectories are time reversible (see \cite{pt18} for a recent generalization). Here we use the more flexible approach of \cite{ka07}, which allows generalization for almost any regions, dimensions, and dynamics.

In this approach, the $d$-dimensional continuous stochastic process $\mathbf X(t)$ takes all its possible initial values inside region $R_0\subseteq \mathbb R^d$ and has an eventual forbidden region $R$, disjoint to $R_0$, whose surface $\partial R=\partial R_r\cup\partial R_a$ is divided into a reflecting $\partial R_r$ part and an absorbing $\partial R_a$ part. The probability density $p(\mathbf x,t)$ of $\mathbf X(t)$ verifies
\begin{eqnarray}
 \label{eq:8}
 && \partial_tp(\mathbf x,t)=-\nabla\cdot \textbf{J}[\mathbf x,t|p], \quad \text{for} \mathbf x\in \mathbb R^d,\\
 \nonumber 
 && p(\mathbf x,t_0)= 0 \quad \text{for} \quad \mathbf x \notin R_0, \int d\mathbf x \ p(\mathbf x, t_0)=1, \\
 \label{eq:9}
 && \mathbf J[\mathbf x,t|p]\cdot \boldsymbol{\hat n}(\mathbf x)=0 \quad \text{for} \quad \mathbf x\in\partial R_r,\\ 
 \nonumber 
 && p(\mathbf x,t)=0 \quad \text{for} \quad \mathbf x\in \partial R_a,
\end{eqnarray}
where $\mathbf J$ denotes the vector of probability flow, which we take as general but is assumed to be a linear functional of $p(\mathbf x,t)$, and $\hat{\boldsymbol{n}}(\mathbf x)$ is a unit vector normal to $\partial R_r$ at point $\mathbf x$ pointing towards the allowed region. We are beyond standard approaches that use a particular dynamics. This way, $p(\mathbf x,t)$ is associated with the ensemble of trajectories that evolve according to the dynamics of \eqref{eq:8}, that begin initially at region $R_0$ with probability density $p(\mathbf x,t_0)$, being rebounded upon arriving at $\partial R_r$, and that are absorbed upon contacting $\partial R_a$.

Once the dynamics of the problem is defined, we compute next the first-passage distribution $f(R,t|R_0,t_0)$ from a region $R_0$ to $R$, with $R$ being a generic $d$-dimensional region disjoint to $R_0$. First, it is useful to divide the trajectories associated with $\mathbf X(t)$ into two sets: the set $S_{\bar R}$ of the trajectories that have never been to region $R$, and the set $S_R$ of the reminder trajectories. The two defined sets are disjoint, so the probability density $p(\mathbf x)$ can be written as
\begin{equation}
 \label{eq:10}
 p(\mathbf x,t)=p_{\bar R}(\mathbf x,t)+p_R(\mathbf x,t),
\end{equation}
where the two terms on the right-hand side account for the contributions of $S_{\bar R}$ and $S_R$, respectively. Second, since the system \eqref{eq:8} and \eqref{eq:9} is a linear problem in $p$, the two new functions are also solutions of Eq.~\eqref{eq:8}, with the same initial condition but different boundary conditions. In particular, it is $p_{\bar R}(\mathbf x,t)=0$ if $\mathbf x\in \partial R$. Finally, we can express $f(R,t|R_0,t_0)$ as a function of $p_{\bar R}(\mathbf x,t)$ as follows. The fraction of trajectories that have never been to $R$ between $t_0$ and $t$ is $\int d\mathbf x \ p_{\bar R}(\mathbf x,t)$, so $-\frac{d}{dt}\int d\mathbf x \ p_{\bar R}(\mathbf x,t)$ gives the rate of loss of probability due to contacts with $R_a$ and $R$, or the fraction of trajectories that have contacted $R_a$ and $R$ for the first time in a time in $(t,t+dt)$, that is, $f(R\cup R_a,t|R_0,t_0)$. After using the equation for $p_{\bar R}$ and the divergence theorem we have
\begin{equation}
 \label{eq:11}
 \begin{split}
 -\frac{d}{dt}\int d\mathbf x \ p_{\bar R}(\mathbf x,t)=&-\int_{\partial R_a}dS\ \mathbf J[\mathbf x,t|p_{\bar R}]\cdot \boldsymbol{\hat n}(\mathbf x)\\ &-\int_{\partial R}dS\ \mathbf J[\mathbf x,t|p_{\bar R}]\cdot \boldsymbol{\hat n}(\mathbf x),
 \end{split}
\end{equation}
which allows us to make the identification
\begin{equation}
 \label{eq:12}
 f(R,t|R_0,t_0)=-\int_{\partial R}dS\ \mathbf J[\mathbf x,t|p_{\bar R}]\cdot \boldsymbol{\hat n}(\mathbf x),
\end{equation}
from which we also infer
\begin{equation}
 \label{eq:13}
 f(\mathbf x,R,t|\mathbf x_0,t_0)=-\mathbf{J}[\mathbf x,t|p_{\bar R}] \cdot \boldsymbol{\hat n}(\mathbf x), \quad \mathbf x\in \partial R,
\end{equation}
provided $p_{\bar R}(\mathbf x,t_0)=\delta(\mathbf x-\mathbf x_0)$, where $ \delta(\mathbf x) $ is the Dirac delta function.

In summary, in order to obtain the first-passage distributions of a stochastic process $\mathbf X(t)$ we have to solve the original problem for its probability distribution $p_{\bar R}$, with the same initial condition, but imposing absorbing boundary conditions to the region where the first passage occurs, and finally apply Eq.~\eqref{eq:13}. The conclusion, which generalizes the results of \cite{ka07}, holds for general dynamics and was used in Sec.~\ref{sec:2} for the case of the 1D Fokker-Planck equation. Observe that the fundamental relation \eqref{eq:13} can be easily generalized to discrete time and/or space.

\subsection*{The one-dimensional Fokker-Planck case}

The general analysis of the preceding subsection can be particularized for the one-dimensional case, under the assumption that the dynamics is described in the form of the Fokker-Planck equation. In this case, the operator $J$ is given by Eq.~\eqref{eq:14},
\begin{eqnarray}
  \nonumber
  J[x,t|p]=&&A(x)p-\frac{1}{2}\frac{\partial[B(x)p]}{\partial x}\\=&&A(x)p-\frac{1}{2}B(x)\frac{\partial p}{\partial x}-\frac{1}{2}p\frac{\partial[B(x)]}{\partial x},
\end{eqnarray}
and the unit vector $\boldsymbol{\hat n}(\mathbf x)$ reduces to $-1$ or $1$. For the case of bounded $\Delta(x)$ [when $A(x)$ is bounded and $B(x)$ is nonzero], upon using the latter expression for $J$ in Eq.~\eqref{eq:13}, only the term proportional to $\frac{\partial p}{\partial x}$ survives, and we recover Eq.~\eqref{eq:20}. 

\section{Obtaining the integral expression of the first-passage distribution\label{app:1_2}}

The goal of this appendix is to show the steps and the approximations needed to go from the discrete version of the first-passage distribution, Eq.~\eqref{eq:24}, to its integral version, Eq.~\eqref{eq:25}, in the case of a regular $\Delta(x)$. Defining $s_n= \sqrt{\lambda_n t}$, the sum in Eq.~\eqref{eq:24} can be written as a sum over $ s_n $,
\begin{equation}
 \label{eq:24b}
\sum_{s_n} \frac{s_n}{\sqrt{t}} g\left(\frac{y_0}{\sqrt{t}}s_n,\boldsymbol{\Delta}_c\,t/s_n^2\right)\overset{\textbf{.}}{g}(0,\boldsymbol{\Delta}_c\,t/s_n^2)e^{-s_n^2},
\end{equation}
which has the form $(1/\sqrt{t})\sum_{n}h(s_n)$, after an evident identification of $h(s_n)$. By means of dimensional analysis, we can write 
\begin{equation}
\label{eq:a2}
s_{n+1}-s_n = \delta_n \frac{\sqrt{t}}{y_*},
\end{equation}
with $\delta_n$ a dimensionless function. Thus, the sum of Eq.~\eqref{eq:24b} can be written as a Riemann sum 
\begin{equation}
\frac{y_*}{t}\sum_{n}h(s_n)\frac{s_{n+1}-s_n}{\delta_n}.
\end{equation}
According to the classical spectral theorem \cite{resi75} applied to Eq.~\eqref{eq:28b}, the eigenvalues satisfy 
\begin{equation}
\label{eq:a1}
n^2+m\le k y_*^2 \lambda_n \le n^2+M,
\end{equation}
where $m$ and and $M$ are bounds such as $m\le k y_*^2 |\Delta| \le M$, $k$ is a constant of order one, and $y_*$ is defined in Eq.~\eqref{eq:28e}. The latter relation implies $s_{n+1}-s_n\sim \frac{\sqrt{t}}{y_*}$, or equivalently $\delta_n\sim 1$, for $n\gg \sqrt{M}$. In other words, for large enough $ n $ the eigenvalues grow quadratically with $ n $ and the interval $s_{n+1}-s_n$ becomes independent of $ n $.

At this point, our first approximation is to take $\delta_n\sim 1$ as a constant, that is to (significantly) modify the contribution to Eq.~\eqref{eq:24b} of addends with $n\lesssim \sqrt{M}$. This approximation is good as long as time is small enough ($t\lambda_{\sqrt{M}}\lesssim 1$) so the smallest modes ($n\lesssim \sqrt{M}$) are ``inactive''. We can estimate the value of the eigenvalue from which this approximation starts to fail, which is, according to the last inequality in Eq.~\eqref{eq:a1}, $\lambda_{\sqrt{M}}\sim M/y_*^2\sim \Delta_*$, with  $\Delta_* = \max_{x}|\Delta(x)|$. Hence, the first approximation is valid for $t\lesssim \Delta_*^{-1}$. 

Our second approximation assumes $s_{n+1}-s_n$ to be small, so that $|h(s_{n+1})-h(s_{n})|=\mathcal{O}(s_{n+1}-s_n)$. According to Eq.~\eqref{eq:a2} and since $h(s_n)$ is a smooth function for bounded $\Delta$, it is enough to take the effective interval $I$ large and/or the time small $\sqrt{t}\ll y_*$.

Under these two approximations, Eq.~\eqref{eq:24b} coincides with its Riemann integral, and Eq.~\eqref{eq:24} becomes the integral expression Eq.~\eqref{eq:25}.

\section{Exact first-passage and first-return distributions for the random walk\label{app:2}}

In this appendix we give a thorough compilation of analytical results regarding the random walk, namely, the first-passage time distributions of the cases I--IV discussed in the main text (see Fig.~\ref{fig:2a}), and a way of obtaining the first-return distribution that is consistent with the discrete numerical simulations. First, consider the Fokker-Planck equation for the probability density $p(x,t)$ with $A(x)=0$ and constant diffusion coefficient $B(x) \equiv B$, with an initial condition $p(x,0) = \delta(x-x_0) $ and a reflecting condition in one of the boundaries, say, at $ \min I=-1 $, and an absorbing condition at the point $x_f=\max I$ in which we want to compute the first-passage time. Thus, due to the one dimensionality of the problem, the effective interval in which the process takes place is $ x\in[-1,x_f] $, assuming $ -1 < x_f $. 
The scenario $ x\in [x_f,1] $, with $ x_f < 1 $, is equivalent because of the symmetries of the problem. The resulting eigenvalue problem \eqref{eq:17} is easily solved and the probability density reads
\begin{equation}
\begin{split}
 p(x,t) = \frac{2}{x_f + 1} \sum_{n=0}^{\infty}& \cos\left[\lambda_n (x+1) \right] \\ & \times \cos\left[\lambda_n (x_0+1) \right] e^{-\frac{B}{2} \lambda_n^2t},
\end{split}
\end{equation}
where the eigenvalues are
\begin{equation}
 \lambda_n = \frac{\pi(n+\frac12)}{(x_f+1)}, \quad n=0,1,\dots
\end{equation}
Now, by using the relation \eqref{eq:20}, the first-passage distribution is
\begin{equation}
 \label{eq:a3_fpt}
 \begin{split}
 f(x_f,t|x_0) =\frac{B}{x_f+1} \sum_{n=0}^{\infty} &(-1)^n \cos\left[ \lambda_n (x_0+1) \right] \\
 & \times \lambda_n e^{-\frac{B}{2} \lambda_n^2t},
 \end{split}
\end{equation}
which can be also written in terms of the Jacobi $\vartheta_1$-function,
\begin{equation}
\vartheta_1(u,q) = 2\sum_{n=0}^{\infty} (-1)^n q^{(n+1/2)^2} \sin\left[(2n+1)u\right],
\end{equation}
as
\begin{equation}
 \begin{split}
 f(x_f,t|x_0) =\frac{B}{2(x_f+1)} \frac{d}{dx_0} \vartheta_1\left(\frac{\pi(x_0+1)}{2(x_f+1)},e^{-\frac{\pi^2B}{2(x_f+1)^2}t}\right).
 \end{split}
\end{equation}

For the case of the first-return distributions, the fact that the departure and target positions are the same complicates the calculations. The problem is ill-defined, in the sense that we cannot impose simultaneously absorbing and reflecting boundary conditions at the same point. 
If we take $x_0=x_f-\epsilon$, then
\begin{equation}
 \begin{split}
 \cos \left[\lambda_n(x_0+1)\right]&=\cos\left[\frac{x_f+1-\epsilon}{x_f+1}\pi\left(n+\frac12\right)\right] \\ &
 =(-1)^n\lambda_n \epsilon+O([(n+1/2)\epsilon]^2),
 \end{split}
\end{equation}
which is a good approximation if $n\epsilon$ is small, or equivalently if $n<n_M\equiv1/\epsilon$. However, if we restrict ourselves to values of the time $t$ bigger than $t_m\sim \frac{2}{B}\frac{(x_f+1)^2}{\pi^2(n_M+1/2)^2}$, then the contribution of most of the addends for $n>n_M$ to the sum on \eqref{eq:a3_fpt} is negligible, since the exponential decay dominates the prefactors. Hence, for $t>t_m$,
\begin{equation}
\label{eq:fret}
 \begin{split}
 f(x_f,t|x_0) \simeq \frac{B}{x_f+1}\epsilon \sum_{n=0}^{\infty} \lambda_n^2 e^{-\frac{B}{2} \lambda_n^2t}.
 \end{split}
\end{equation}
Using the Jacobi $\vartheta_2$-function, defined as
\begin{equation}
\vartheta_2(u,q) = 2 \sum_{n=0}^{\infty} q^{(n+1/2)^2} \cos\left[(2n+1)u\right],
\end{equation}
the return distribution is
\begin{equation}
 \label{eq:res}
 f(x_f,t|x_0)\simeq -\frac{\epsilon}{x_f+1} \frac{d}{dt}\vartheta_2\left(0,e^{-\frac{\pi^2B}{2(x_f+1)^2}t}\right).
\end{equation}
The latter function behaves like $t^{-3/2}$ for small times which implies that we cannot take $\epsilon\to 0$ naively, since we then have $n_M\to \infty$ and $t_m\to 0$ but the resulting function \eqref{eq:res} has an infinite norm. To overcome this problem, we can just take $\frac{1}{N}$ as the minimal time, the time step of the discrete model studied by means of Monte Carlo simulations, and approximate the return distribution as the normalization function of the latter expression, namely
\begin{equation}
f(x_0,t|x_0)\simeq \frac{-1}{\vartheta_2\left(0,e^{-\frac{\pi^2B}{2N(x_f+1)^2}}\right)}\frac{d}{dt}\vartheta_2\left(0,e^{-\frac{\pi^2B}{2(x_f+1)^2}t}\right),
\end{equation}
for $t>1/N$ and zero otherwise.

\end{document}